\documentclass[aps,showpacs,preprintnumbers,nofootinbibt,twocolumn]{revtex4}
\usepackage{amssymb}
\usepackage{graphicx}
\usepackage{color}

\def\be{\begin{equation}}
\def\ee{\end{equation}}
\def\bea{\begin{eqnarray}}
\def\eea{\end{eqnarray}}

\begin{document}

\title{Bose-Einstein condensate strings}
\author{Tiberiu Harko$^{1}$}
\email{t.harko@ucl.ac.uk}
\author{Matthew J. Lake$^{2,3}$}
\email{matthewj@nu.ac.th}
\affiliation{$^1$Department of Mathematics, University College London, Gower Street,
London, WC1E 6BT, United Kingdom}
\affiliation{$^2$ The Institute for Fundamental Study, ``The Tah Poe Academia Institute",
\\
Naresuan University, Phitsanulok 65000, Thailand and \\
$^3$ Thailand Center of Excellence in Physics, Ministry of Education,
Bangkok 10400, Thailand }

\begin{abstract}
We consider the possible existence of gravitationally bound general
relativistic strings consisting of Bose-Einstein condensate (BEC) matter
which is described, in the Newtonian limit, by the zero temperature
time-dependent nonlinear Schr\"odinger equation (the Gross-Pitaevskii
equation), with repulsive interparticle interactions. In the Madelung
representation of the wave function, the quantum dynamics of the condensate
can be formulated in terms of the classical continuity equation and the
hydrodynamic Euler equations. In the case of a condensate with quartic
nonlinearity, the condensates can be described as a gas with two pressure
terms, the interaction pressure, which is proportional to the square of the
matter density, and the quantum pressure, which is without any classical
analogue though, when the number of particles in the system is high enough,
the latter may be neglected. Assuming cylindrical symmetry, we analyze
the physical properties of the BEC strings in both the interaction pressure
and quantum pressure dominated limits, by numerically integrating the
gravitational field equations. In this way we obtain a large class of stable
stringlike astrophysical objects, whose basic parameters (mass density and
radius) depend sensitively on the mass and scattering length of the
condensate particle, as well as on the quantum pressure of the Bose-Einstein
gas.
\end{abstract}

\maketitle




\section{Introduction}

\label{Sect.I} The observation, in 1995, of Bose-Einstein condensation in
dilute alkali gases, such as vapors of rubidium and sodium, confined in a
magnetic trap and cooled to very low temperatures \cite{exp}, represented a
major breakthrough in experimental condensed matter physics, as well as a
major confirmation of an important prediction in theoretical statistical
physics. At very low temperatures, all particles in a dilute Bose gas
condense to the same quantum ground state, forming a Bose-Einstein
condensate (BEC), which appears as a sharp peak over a broader distribution
in both coordinate and momentum space. Particles become correlated with each
other when their wavelengths overlap, that is, when the thermal wavelength $%
\lambda _{T}$ is greater than the mean interparticle distance $l$. This
happens at a critical temperature $T_c<2\pi \hbar ^{2}n^{2/3}/mk_{B}$, where
$m$ is the mass of an individual condensate particle, $n$ is the number
density, and $k_{B}$ is Boltzmann's constant \cite{Da99,
rev,Ch05,Pit,Pet,Zar}. A coherent state develops when the particle density
is high enough, or the temperature is sufficiently low. From an experimental
point of view, the condensation is indicated by the generation of a sharp
peak in the velocity distribution, which is observed below the critical
temperature \cite{exp}. More recently, quantum degenerate gases have been created by a
combination of laser and evaporative cooling techniques, opening several new
lines of research at the border of atomic, statistical and condensed matter
physics \cite{Da99,rev,Ch05,Pit,Pet,Zar}.

Since Bose-Einstein condensation is a phenomenon that has been observed and
well studied in the laboratory, the possibility that it may occur on
astrophysical or cosmic scales cannot be rejected \textit{a priori}. Thus,
dark matter, which is required to explain the dynamics of the neutral
hydrogen clouds at large distances from the Galactic center, and which is a
cold bosonic system, could also exist as a Bose-Einstein condensate \cite%
{Sin}. In fact, since there exists a formal analogy between classical scalar
fields and BECs, any theory of scalar field dark matter may also be viewed
as condensate system \cite{SF-BEC}. In early studies, such as those given in
\cite{Sin}, either a phenomenological approach was adopted, or the solutions
of the Gross-Pitaevskii equation, which describes the condensate in the
nonrelativistic limit, were investigated numerically. A systematic study of
the properties of condensed galactic dark matter halos was performed in \cite%
{BoHa07} and these systems have been further investigated by numerous
authors \cite{inv}.

By introducing the Madelung representation of the wave function, the
dynamics of the dark matter halo can be formulated in terms of the
continuity equation and the hydrodynamic Euler equations. Hence, condensed
dark matter can be described as a Newtonian gas, whose density and pressure
are related by a barotropic equation of state. However, in the case of a
condensate with quartic nonlinearity, the equation of state is polytropic
with index $n=1$ \cite{BoHa07}.

Furthermore, though superfluids, such as liquid $^{4}$He, are far from being
dilute, there is, nevertheless, good reason to believe that the phenomenon
of superfluidity is related to that of Bose-Einstein condensation. Both
experimental observations and theoretical calculations estimate the
condensate fraction at $T=0$ (denoted $n_0$) for superfluid helium to be
around $n_{0}\approx 0.10$ and, since a strongly correlated pair of fermions
can be treated approximately like a boson, the arising superfluidity
can be interpreted as the condensation of coupled fermions. Similarly, the
transition to a superconducting state in a solid material may be described
as the condensation of electrons or holes into Cooper pairs, which
drastically reduces the friction caused by the flow of current \cite{Ch05}.

The possibility of Bose-Einstein condensation in nuclear and quark matter
has also been considered in the framework of the so-called
Bardeen-Cooper-Schrieffer to Bose-Einstein condensate (BCS-BEC) crossover
\cite{BCS-BEC}. At ultrahigh density, matter is expected to form a
degenerate Fermi gas of quarks in which Cooper pairs of quarks form a
condensate near the Fermi surface (a color superconductor \cite{csc}). If
the attractive interaction is strong enough, at some critical temperature
the fermions may condense into the bosonic zero mode, forming a quark BEC
\cite{Bal95}. In general, fermions exist in a BCS state when the attractive
interaction between particles is weak. The system then exhibits
superfluidity, characterized by the energy gap for single-particle
excitations which is created by the formation of the Cooper pairs.
Conversely, a BEC exists when the attractive interaction between fermions is
strong, causing them to first form bound ``molecules" (i.e. bosons), before
starting to condense into the bosonic zero mode at some critical
temperature. An important point is that the BCS and BEC states are smoothly
connected, without a phase transition between the two \citep{NiAb05} 
(see also \cite{Sun:2007fc,He:2007kd,He:2007yj,He:2010nb,He:2013gga} 
for additional work on the BCS-BEC crossover in relativistic matter).

Remarkably, the critical temperature in the BEC region is, in fact,
independent of the precise strength of the coupling for the attractive
interaction between fermions. This is because an increase in the coupling
strength affects only the internal structure of the bosons, whereas the
critical temperature is determined by their kinetic energy. Thus, the
critical temperature reaches an upper limit for strong coupling, as long as
the effect of the binding energy on the total mass of the boson is small,
and can be neglected.
In this limit, we are able to use a nonrelativistic framework to describe
the BCS-BEC crossover \citep{NiAb05}.

However, in relativistic systems, the binding energy makes a significant
contribution to the total mass of the boson and cannot be neglected. In this
case, two crossovers are possible. First, an ordinary BCS-BEC crossover
may occur, though the critical temperature in the BEC region no longer tends
towards an upper bound, due to relativistic effects. Second, the
nonrelativistic BEC state undergoes a transition to a relativisitc
BEC (RBEC) state, in which the critical temperature increases to the
order of the Fermi energy \citep{NiAb05}.

The possibility of Bose condensates existing in neutron stars has been
considered (see Glendenning \cite{Gl00} for a detailed discussion), as the
condensation of negatively charged mesons in neutron star matter is favored,
since these mesons would replace electrons with very high Fermi momenta.
Recently, Bose-Einstein condensates of kaons/anti-kaons in compact objects
were also discussed \cite{BaBa03,Ban04}. Pion as well as kaon condensates
would have two important effects on neutron stars. First, condensates
soften the equation of state above the critical density for the onset of
condensation, which reduces the maximal possible neutron star mass. At the
same time, however, the central stellar density increases, due to the
softening. Second, meson condensates would lead to neutrino luminosities
which are considerably enhanced over those of normal neutron star matter.
This would speed up neutron star cooling considerably \citep{Gl00}. Another
particle which may form a condensate is the H-dibaryon, a doubly strange
six-quark composite with zero spin and isospin, and baryon number $B=2$. In
neutron star matter, which may contain a significant fraction of $\Lambda $
hyperons (i.e. neutral subatomic hadrons consisting of one up, one down and
one strange quark, labeled $\Lambda^{0}$) \cite{LambdaHyperon}, these
particles could combine to form H-dibaryons \cite{Hdibar}. Thus, H-matter
condensates may thus exist at the center of neutron stars \citep{Gl00}.
Neutrino superfluidity, as suggested by Kapusta \cite{Ka04}, may also lead
to Bose-Einstein condensation \citep{Ab06}.

These results show that the possibility of the existence of Bose-Einstein
condensed matter inside compact astrophysical objects, or even the existence
of stars formed entirely from a BEC, cannot be excluded \textit{a priori}.
The properties of BEC stars have been considered in \cite{ChHa}, and it was
shown that these hypothetical astrophysical objects have mass and radii
ranges that are compatible with the observed physical parameters of some
neutron stars. Bearing this in mind, it is the purpose of the current paper
to consider another possible astrophysical BEC system, with potentially
important cosmological implications, namely, the Bose-Einstein condensate
string. By this, we mean a cylindrically symmetric system consisting of
bosonic matter in a Bose-Einstein condensed phase.

The structure of this paper is then as follows. In Sec.~\ref{Sect.II}, we
briefly review the general treatment of gravitationally bound Bose-Einstein
condensates in the nonrelativistic limit, including their description by
the generalized Gross-Pitkaevskii equation (Sec.~\ref{Sect.IIA}) and the
hydrodynamical representation (Sec.~\ref{Sect.IIB}) in which a quantum
potential term, which is significant close to the boundary of the
condensate, arises. Using the nonrelativistic analysis as a guide, we
determine that the thermodynamic (``interaction") pressure of BEC dark
matter is governed by a polytropic equation of state which, together with
the assumption of cylindrical symmetry, allows us to fix the general form of
the metric, and of the components of the energy-momentum tensor, which are
expressed in terms of the energy density and pressure of the fluid. The
latter is, in turn, decomposed into interaction and quantum pressure
terms, denoted $p$ and $p_Q$, respectively, and the two limiting cases $p
\gg p_Q$ and $p_Q \gg p$ are considered separately. Thus, the string
geometry and the corresponding form of the Einstein field equations are
determined in Sec.~\ref{Sect.III} and, by appropriately redefining the
relevant dynamical variables, these are then recast (in each limiting case)
as an autonomous system of differential equations, which may be solved
numerically. Specifically, Sec.~\ref{Sect.IIIA} reviews the
semiclassical approximation for the relativistic treatment of the
quantum pressure term and Sec.~\ref{Sect.IIIB} deals with the string
geometry and components of Einstein tensor, while the numerical solutions of
the field equations for interaction dominated and quantum pressure dominated
strings are given in Secs. \ref{Sect.IIIC} and \ref{Sect.IIID},
respectively. The meaning and significance of the variation, with respect to
the radial coordinate $r$, of the physical parameters of the BEC string
(i.e. the three-dimensional energy density and pressure, and the mass per
unit length) and of the components of the metric tensor that determine the
resulting space-time, is discussed, in each limiting case, at the end of the
relevant subsection. In Sec. \ref{new} we carefully consider the effects of the geometry of the space-time on the quantum pressure term. The physical parameters of the string are obtained by numerically integrating the field equations for a fixed set of initial conditions, and for different values of the single free model parameter. The comparison of the two quantum string models is considered and, by analyzing and discussing the behavior of the physical and geometrical quantities, we estimate the effect of the geometry on the global properties of the string.
Finally, Sec.~\ref{Sect.IV} contains a brief
treatment of BEC strings in the Newtonian approximation of the gravitational
field, from which it is seen that constraints on the order of magnitude
values of physically important quantities can be easily obtained. A summary
of the results for both the thermodynamic and quantum pressure dominated
regimes, together with some brief remarks regarding their possible
cosmological and astrophysical significance, and suggestions for future
work, are given in Sec.~\ref{Sect.V}.

\section{Bose-Einstein Condensation}

\label{Sect.II}

In a quantum system of $N$ interacting condensed bosons, most of the bosons
lie in the same single-particle quantum state. For a system consisting of a
large number of particles, the calculation of the ground state of the system
with the direct use of the Hamiltonian is impracticable, due to the high
computational cost. However, the use of some approximate methods can lead to
a significant simplification of the formalism. One such approach is the mean
field description of the condensate, which is based on the idea of
separating out the condensate contribution to the bosonic field operator. We
also assume that, in a medium composed of scalar particles with nonzero
mass, when the transition to a Bose-Einstein condensed phase occurs, the
range of Van der Waals-type scalar mediated interactions among particles
becomes infinite.


\subsection{The  Gross-Pitaevskii equation}

\label{Sect.IIA}

The many-body Hamiltonian describing interacting bosons confined by an
external potential $V_{ext}$ is given, in the second quantization, by
\begin{eqnarray}  \label{ham}
&&\hat{H}=\int d\vec{r}\hat{\Phi}^{+}\left( \vec{r}\right) \left[ -\frac{%
\hbar ^{2}}{2m}\vec{\nabla} ^{2}+V_{ext}\left( \vec{r}\right) \right] \hat{%
\Phi}\left( \vec{r}\right) +  \nonumber \\
&&\frac{1}{2}\int d\vec{r}d\vec{r}^{\prime }\hat{\Phi}^{+}\left( \vec{r}%
\right) \hat{\Phi}^{+}\left( \vec{r}^{\prime }\right) V_{int}\left( \vec{r}-%
\vec{r}^{\prime }\right) \hat{\Phi}\left( \vec{r}\right) \hat{\Phi}\left(
\vec{r}^{\prime }\right),
\end{eqnarray}
where $\vec{\nabla}$ is the three-dimensional Laplacian, $\hat{\Phi}\left(
\vec{r}\right) $ and $\hat{\Phi}^{+}\left( \vec{r}\right) $ are the boson
field operators that annihilate and create a particle at the position $\vec{r%
}$, respectively, and $V_{int}\left( \vec{r}-\vec{r}^{\prime }\right) $ is
the two-body interatomic potential \cite{Da99, Ch05,Pet}. $V_{ext}\left(
\vec{r}\right)$ represents an ``externally applied" potential. However, in
the nonrelativistic limit, Newtonian gravitational potentials, even those
corresponding to the gravitational field of the condensate itself, may be
considered as externally applied potentials in this sense.

In order to simplify this formalism, we now adopt the mean field
approximation and separate out the contribution to the bosonic field
operator. For a uniform gas in a volume $\mathcal{V}$, a BEC forms in the
single-particle state $\Phi _{0}=1\sqrt{\mathcal{V}} $, having zero
momentum. The field operator can then be decomposed as $\hat{\Phi}\left(
\vec{r}\right) =\sqrt{\mathcal{N}/\mathcal{V}}+\hat{\Phi}^{\prime }\left(\vec{r}%
\right) $, where $\mathcal{N}$ is the total number of particles in the condensate. By treating the operator $\hat{\Phi}^{\prime }\left( \vec{r}%
\right) $ as a small perturbation, one can develop the first-order theory
for the excitations of the interacting Bose gases \cite{Da99,Ba01}.

In the general case of a nonuniform and time-dependent configuration the
field operator in the Heisenberg representation is given by $\hat{\Phi}%
\left( \vec{r},t\right) =\psi \left( \vec{r},t\right) +\hat{\Phi}^{\prime
}\left( \vec{r},t\right) $, where $\psi \left( \vec{r},t\right) $, also
called the condensate wave function, is the expectation value of the field
operator, $\psi \left( \vec{r},t\right) =\left\langle \Psi|\hat{\Phi}\left( \vec{r%
},t\right)|\Psi  \right\rangle $. It is a classical field, and its absolute value
fixes the number density of the condensate particles according to $\rho_{\mathcal{N}}
\left( \vec{r},t\right) =\left| \psi \left( \vec{r},t\right) \right| ^{2}$.
The normalization condition is $\mathcal{N}=\int \rho_{\mathcal{N}} \left( \vec{r},t\right) d^{3}%
\vec{r}$.

The equation of motion for the condensate wave function is the Heisenberg
equation corresponding to the many-body Hamiltonian given by Eq.~(\ref{ham}%
),
\begin{eqnarray}
&&i\hbar \frac{\partial }{\partial t}\hat{\Phi}\left( \vec{r},t\right) =%
\left[ \hat{\Phi},\hat{H}\right] = \left[ -\frac{\hbar ^{2}}{2m}\vec{\nabla}
^{2}+V_{ext}\left( \vec{r}\right)\right. +  \nonumber \\
&&\left.\int d\vec{r}^{\prime }\hat{\Phi}^{+}\left( \vec{r}^{\prime
},t\right) V_{int}\left( \vec{r}^{\prime }-\vec{r}\right) \hat{\Phi}\left(
\vec{r}^{\prime },t\right) \right] \hat{\Phi}\left( \vec{r},t\right).
\label{gp}
\end{eqnarray}

The zeroth-order approximation to the Heisenberg equation is obtained by
replacing $\hat{\Phi}\left( \vec{r},t\right) $ with the condensate wave
function $\psi\left( \vec{r},t\right) $. In the integral containing the
particle-particle interaction, $V_{int}\left( \vec{r}^{\prime }-\vec{r}%
\right)$, this replacement is, in general, a poor approximation for short
distances. However, in a dilute and cold gas, only binary collisions at low
energy are relevant, and these collisions are characterized by a single
parameter, the $s$-wave scattering length $l_s$, independently of the
details of the two-body potential. Therefore, one can replace $V_{int}\left(
\vec{r}^{\prime }-\vec{r}\right) $ with an effective interaction $%
V_{int}\left( \vec{r}^{\prime }-\vec{r}\right) =u_0 \delta \left( \vec{r}%
^{\prime }-\vec{r}\right) $, where  $u_0 =4\pi \hbar ^{2}l_s/m$,  $l_s$ is the coherent scattering length
(defined as the zero-energy limit of the scattering amplitude, $l_s =
\lim_{T \rightarrow 0}f_{scat}$, and  $m$ is the mass of an individual condensate particle. With the use of
the effective potential the integral in the bracket of Eq.~(\ref{gp}) gives $%
u_0 \left| \psi \left( \vec{r},t\right) \right| ^{2}$, and the resulting
equation is the Schr{\" o}dinger equation with a quartic nonlinear term %
\cite{Da99,rev,Pit, Pet, Zar, BoHa07}. This gives the generalized Gross-Pitaevskii
equation, describing a gravitationally trapped  Bose-Einstein
condensate in the nonrelativistic limit,
\begin{eqnarray}  \label{sch_1}
i\hbar \frac{\partial }{\partial t}\psi \left( \vec{r},t\right) &=&\bigg[ -%
\frac{\hbar ^{2}}{2m}\vec{\nabla} ^{2}+V_{ext}\left( \vec{r}\right) +
\nonumber \\
&&u_0 \left| \psi \left( \vec{r},t\right) \right| ^{2} \bigg] \psi
\left( \vec{r},t\right) .
\end{eqnarray}
As for $V_{ext}\left( \vec{r}\right) $, we assume that it is the Newtonian
gravitational potential, which we denote as $V_{grav}$, 
which satisfies the Poisson equation
\begin{equation}
\vec{\nabla} ^{2}V_{grav}=4\pi G\rho ,
\end{equation}
where $\rho =m\rho_{\mathcal{N}} =m\left| \psi \left( \vec{r},t\right) \right| ^{2}$ is
the mass density inside the 
condensate.


\subsection{The hydrodynamical representation}

\label{Sect.IIB}

The physical properties of a Bose-Einstein condensate described by the
generalized Gross-Pitaevskii equation, Eq.~(\ref{sch_1}), can be understood
much more easily using the so-called Madelung representation of the wave
function \cite{Da99}, which here consists of writing $\psi\left( \vec{r}%
,t\right)$ in the form
\begin{equation}
\psi \left( \vec{r},t\right) =\sqrt{\frac{\rho \left( \vec{r},t\right)}{m} }%
\exp \left[ \frac{i}{\hbar }S\left( \vec{r},t\right) \right] ,
\end{equation}
where the function $S\left( \vec{r},t\right) $ has the dimensions of an
action. Substituting the above expression for $\psi \left( \vec{r},t\right) $
into Eq.~(\ref{sch_1}), the latter decouples into a system of two
differential equations for the real functions $\rho $ and $\vec{v}$, given
by
\begin{equation}
\frac{\partial \rho }{\partial t}+\vec{\nabla} \cdot \left( \rho \vec{v}%
\right) =0,  \label{cont}
\end{equation}
\begin{eqnarray}  \label{euler}
\rho \left[ \frac{\partial \vec{v}}{\partial t}+\left( \vec{v}\cdot \vec{%
\nabla} \right) \vec{v}\right] &=&-\vec{\nabla} P\left( \frac{\rho }{m}%
\right)- \rho \vec{\nabla} \left( \frac{V_{ext}}{m}\right) -  
\rho \vec{\nabla} V_{Q}, \nonumber \\&&
\end{eqnarray}
where we have introduced the quantum potential
\begin{equation}
V_{Q}=-\frac{\hbar ^{2}}{2m}\frac{\vec{\nabla} ^{2}\sqrt{\rho }}{\sqrt{\rho }%
},
\end{equation}
the velocity of the quantum fluid, defined as
\begin{equation}
\vec{v}=\frac{\vec{\nabla} S}{m},
\end{equation}
and have denoted
\begin{equation}
P\left( \rho \right) =U_{0}\rho ^{2},
\end{equation}
where
\begin{eqnarray}
U_{0}&=&\frac{2\pi \hbar ^{2}l_s}{m^{3}}=  \nonumber \\
&&1.232\times10^{50}\left(\frac{m}{1\;\mathrm{meV}}\right)^{-3}\left(\frac{%
l_s}{10^9\;\mathrm{fm}}\right)\;\mathrm{cm}^5/\mathrm{g}\;\mathrm{s}^2,
\nonumber \\
\end{eqnarray}
or
\begin{equation}
U_0
=0.1856\times 10^5 \left(\frac{l_s}{1\;%
\mathrm{fm}}\right)\left(\frac{m}{2m_n}\right)^{-3},
\end{equation}
where $m_n=1.6749\times 10^{-24}$ g is the mass of the neutron.

Therefore, the equation of state of the Bose-Einstein condensate with
quartic nonlinearity is a polytrope with index $n=1$. However, in the case
of low-dimensional systems \cite{Kolo} it has been shown that, in many
experimentally interesting cases, the nonlinearity will be cubic, or even
logarithmic in $\rho$.

From its definition it follows that the velocity field is irrotational,
satisfying the condition $\vec{\nabla} \times \vec{v}=0$. Therefore the
equations of motion for the gravitationally bound, ideal Bose-Einstein
condensate (in the nonrelativistic limit), take the form of the continuity
equation plus the hydrodynamic Euler equation, with the density and pressure
related by a barotropic equation of state \cite{Da99,Ch05,Pet}.

By taking into account the mathematical identity
\begin{equation}
\tau \vec{\nabla} \left(\frac{\vec{\nabla}^2 \sqrt{\tau}}{\sqrt{\tau}}%
\right)=\frac{1}{2}\vec{\nabla} \left(\tau \vec{\nabla}^2 \ln \tau\right),
\end{equation}
which holds for any dimensionless scalar function $\tau$, it follows
that the quantum potential $V_Q$ generates a quantum pressure $p_Q$, given
in a general form as \cite{Da99,Ch05,Pet}
\begin{equation}  \label{quantpress}
p_Q = \frac{\rho \vec{\nabla}V_Q}{\hbar^2} = -\frac{\hbar ^2}{4m^2}\rho \vec{%
\nabla}^2 \ln (\rho/\rho_c),
\end{equation}
where $\rho_c$ is the central density of the BEC mass distribution.
This pressure can have a significant effect for small particle masses and
high densities.

When the number of particles becomes large enough, the quantum pressure term makes a significant
contribution only near the boundary of the condensate. Hence it is much
smaller than the nonlinear interaction term. Thus, the quantum stress term
in the equation of motion for the condensate can be neglected. This is the
Thomas-Fermi approximation, which has been used extensively for the study of
Bose-Einstein condensates \cite{Da99,Ch05,Pet}. As the number of
particles in the condensate becomes infinite, the Thomas-Fermi approximation
becomes exact. This approximation also corresponds to the classical limit of
the theory, i.e. to neglecting all terms in nonzero powers of $\hbar$, or,
equivalently, to the regime of strong repulsive interactions among
particles. In the hydrodynamical representation, the Thomas-Fermi
approximation corresponds to neglecting all terms containing $\vec{\nabla} {%
\rho }$ (or $\vec{\nabla} \sqrt{\rho}$) and $\vec{\nabla} {S}$ in the
equations of motion.

For cylindrical distributions, such as those considered (relativistically)
in the remainder of this study, the corresponding equation of motion in the
nonrelativistic limit is the ``cylindrical Lane-Emden equation", $%
\left(1/\xi \right)d\left(\xi d\theta /d\xi\right)/d\xi +\theta ^n=0$, which
is equivalent to $\left(1/\xi \right)d\left((\xi/n) (d\tau /d\xi) \tau^{\frac{1}{%
n}-1}\right)/d\xi + \tau=0$, where $\theta^n = \tau = \rho/\rho_{c}$, for finite $n$,
where $\rho _c$ is the central density at $r=0$. For a barotropic fluid, the
limit $n \rightarrow \infty$ must be taken before substituting the
equation of state into the Poisson equation, leading to the separate
(critical) case, for which $\left(1/\xi \right)d\left(\xi d(\ln(\tau))
/d\tau \right)/d\xi + \tau=0$. In this case, we have that $\Gamma
\rightarrow 2$, $\alpha \rightarrow U_0$ and $\xi = [U_0/(4\pi
G\rho_{c})]^{1/2}r$. This equation can also be generalized to include
rotating fluids by adding a term of the form $\Omega^2\xi$, where $\Omega =
\Omega(\xi)$ is a function of the rescaled radial coordinate $\xi$ only, to
the right-hand side \cite%
{Ostriker(1964),Schneider+Schmitz(1995),Christodoulou+Kazanas(2007)}.\newline
\indent
Hence, all Bose-Einstein condensate forms of matter can generally be
described as fluids satisfying a polytropic equation of state of index $n$
and, importantly, this remains true relativistically, as well as in the
nonrelativistic limit, which, in effect, has been thoroughly studied in
\cite{Ostriker(1964),Schneider+Schmitz(1995),Christodoulou+Kazanas(2007)},
at least in the limit of the Fermi-Thomas approximation. The remainder of
this study is therefore dedicated to determining the astrophysical and
cosmological significance of stringlike (i.e. cylindrically symmetric) BEC
dark matter structures, by studying them in a general relativistic context.
For simplicity, we consider only the case of the condensates with quartic
nonlinearity since, in this case, the physical properties of the condensate
are relatively well known from laboratory experiments and can be described
in terms of only two free parameters, the mass $m$ of the condensate
particle, and the scattering length $l_s$. %

\section{Static Bose-Einstein condensate strings in cylindrically symmetric
geometries}

\label{Sect.III}

In this section we consider the properties of a string consisting of
matter in a Bose-Einstein condensed state. In the hydrodynamical description
of the condensate the equilibrium properties of this system are determined
by two physical parameters, the interaction pressure $p$, and the quantum
pressure $p_Q$, respectively. In the following we will investigate two
classes of BEC strings, corresponding to the conditions $p>>p_Q$
(interaction energy dominated strings), and $p_Q>>p$ (quantum pressure
dominated strings).


\subsection{General relativistic Bose-Einstein condensates - the
semiclassical approximation}

\label{Sect.IIIA}

In formulating our initial, general relativistic, model of gravitationally bound
Bose-Einstein condensates, we consider that bosonic matter at temperatures
below the critical temperature $T_c$ represents a hybrid system, in which
the gravitational field remains classical, while the bosonic condensate is
described by quantum fields in which gravitational effects induced by the
non-Euclidian space-time geometry can be effectively neglected. In the
standard approach used for coupling quantum fields to a classical
gravitational field (i.e. semiclassical gravity), the energy-momentum tensor
that serves as the source in the Einstein equations is replaced by the
expectation value of the energy-momentum operator $\hat{T}_{\mu \nu}$, with
respect to some quantum state $\Psi $ \cite{Carl},
\begin{equation}  \label{gr1}
R_{\mu \nu}-\frac{1}{2}g_{\mu \nu}R=\frac{8\pi G}{c^4}\left \langle\Psi
\left |\hat{T}_{\mu \nu}\right |\Psi \right \rangle,
\end{equation}
where $R_{\mu \nu}$ is the Ricci tensor, $R$ is the scalar curvature, and $%
g_{\mu \nu}$ is the metric tensor of the space-time. In the nonrelativistic
limit, the state function $\Psi $ evolves according to the Gross-Pitaevski
equation and $\Psi \rightarrow \psi$ so that, for a quartic nonlinear term, the evolution of the
condensate wave function $\psi$ is determined by Eq.~(\ref{sch_1}),
with $V_{ext}$  being given by the gravitational potential $V_{grav}$,
which is the Newtonian limit of Eq.~(\ref{gr1}) for a cylindrically symmetric system. Hence, in the
semiclassical approach, we obtain
\begin{equation}
\left \langle\Psi \left |\hat{T}_{\mu \nu}\right |\Psi \right \rangle\approx \left
\langle\psi \left |\hat{T}_{\mu \nu}\right |\psi \right \rangle=T_{\mu \nu},
\end{equation}
as the source term in the Einstein field equations, where $T_{\mu \nu}$ is
the effective energy-momentum tensor of the condensate system obtained from
the Gross-Pitaevskii equation. In a comoving frame
$T_{\mu \nu}$ is diagonal with components
\begin{equation}
T_{\mu \nu}=\left(\rho c^2, -P,-P,-P\right),
\end{equation}
where $\rho c^2$ denotes the three-dimensional energy density and $P$
denotes the total effective thermodynamic pressure of the system,
obtained from the hydrodynamic representation.
That is, $P \approx p + p_Q$, where $p$ is the genuine classical
(interaction) thermodynamic pressure and $p_Q$ is the quantum pressure term,
as discussed previously. Therefore, in the first approximation, we can assume that the
effective thermodynamic properties (energy density and pressure) of the Bose-Einstein condensed matter are given by the relations derived from the
{\it quantum} Gross-Pitaevskii equation in the Newtonian approximation of
the gravitational field and in a standard Euclidian quantum
geometry. However, geometric effects may play an important role in the equation of state of the quantum string, and we will consider the effects these may have on Eq.~(\ref{quantpress}) by appropriately defining the operator $\vec{\nabla }^2$ in the  given Riemannian geometry.


\subsection{Geometry and gravitational field equations of Bose-Einstein
condensate strings}

\label{Sect.IIIB}

For the geometrical description of Bose-Einstein condensate strings we adopt
cylindrical polar coordinates $(x^0=t,x^1=r,x^2=\phi,x^3=z)$ and assume a
cylindrically symmetric metric, which gives rise to a line element of the
form \cite{clv}
\begin{equation}
ds^{2}= g_{\mu\nu}dx^{\mu}dx^{\nu} = N^{2}(r)dt^{2}-d{r}^{2}-L^{2}(r)d{\phi }%
^{2}-K^{2}(r)dz^{2},  \label{lineelement}
\end{equation}%
where $N(r)$, $L(r)$ and $K(r)$ are arbitrary functions of the radial
coordinate $r$. The nonzero Christoffel symbols associated to the metric (%
\ref{lineelement}) are given by
\begin{eqnarray}  \label{Christoffel}
&&\Gamma _{rt}^{t}=\frac{N^{\prime }(r)}{N(r)},\Gamma
_{tt}^{r}=N(r)N^{\prime }(r),\Gamma _{r\phi}^{\phi}=\frac{L^{\prime }(r)}{%
L(r)},  \nonumber \\
&&\Gamma _{\phi\phi}^{r}=-L(r)L^{\prime }(r),\Gamma _{rz}^{z}=\frac{%
K^{\prime }(r)}{K(r)},\Gamma _{zz}^{r}=-K(r)K^{\prime }(r),  \nonumber \\
&&
\end{eqnarray}%
where a prime denotes the derivative with respect to $r$,
and the nonzero components of the Ricci tensor are \cite{clv}
\begin{eqnarray}  \label{Ricci}
R_{t}^{t}&=&\frac{(LKN^{\prime })^{\prime }}{NLK} ,R_{r}^{r}=\frac{N^{\prime
\prime }}{N}+\frac{L^{\prime \prime }}{L}+\frac{K^{\prime \prime }}{K},
\nonumber \\
&&R_{\phi }^{\phi }=\frac{(NKL^{\prime })^{\prime }}{NLK} ,R_{z}^{z}=\frac{%
(NLK^{\prime })^{\prime }}{NLK}.
\end{eqnarray}


\subsection{Interaction energy dominated Bose-Einstein condensate strings}

\label{Sect.IIIC}

In this section, we assume that the quantum pressure term inside the BEC
string is negligible. This corresponds to the Thomas-Fermi approximation,
which is valid when the number of particles is very large. However, in the
context of a cylindrically symmetric distribution, we must remember that the
term large refers to the number of particles present in a thin,
effectively two-dimensional slice of the string, i.e. a perpendicular
cross section, rather than the total number within the string as a whole. As
such, it is unlikely that this assumption will hold with any degree of
accuracy for very narrow strings and, in general, we would expect the ratio
of the string surface area to its internal volume to play a significant role
in its dynamics, especially for narrow stings in which the ``boundary
region" occupies a significant proportion of the overall volume. In other
words, we would expect the internal quantum dynamics of the string to play a
significant role in determining its macroscopic (essentially classical)
dynamics, through the generation of an effective surface tension.

As an immediate corollary, we see that, were we to attempt to derive an
effective action for a BEC string, it would not be possible to simply take
the ``wire approximation" (i.e. the zero thickness limit \cite%
{Anderson,topological_defects}), as used, for example, to derive the
Nambu-Goto action \cite{NambuGoto} as the effective action for
Nielsen-Olesen strings \cite{no}. Rather, we would need to develop a
modified Nambu-type action (such as those corresponding to species of
chiral, superconducting, or current-carrying string, e.g. in \cite%
{NiOl87,BlOlVi01,CoHiTu87}), incorporating surface tension effects, possibly
through the existence of an additional rigidity and/or elasticity terms
(c.f. \cite{Anderson et al.(1997),Gregory(1988),Gregory(1988)*}). Though
such an analysis remains beyond the scope of the current paper, it would
form the next logical step in the study of BEC strings \cite{HarkoLake2014}.
In principle, for strings in which the thermodynamic pressure $p$ and
quantum pressure $p_Q$ are both significant (which are not considered
in the present study), both finite-thickness corrections (c.f. \cite%
{MaedaTurok(1988)}) and surface tension effects may also need to be
incorporated.

However, within the limit of the Thomas-Fermi approximation, the
Bose-Einstein condensate can be described as a quantum gas satisfying a
polytropic equation of state with index $n=1$. Therefore, the source term in
the field equations is given by the energy-momentum tensor of a BEC, with
the following components
\begin{eqnarray}  \label{Tmunu}
T_{t}^{t} = \rho (r)c^2, T_{r}^{r} =T_{\phi }^{\phi } =T_{z}^{z}=-p(r).
\end{eqnarray}
We then have
\begin{eqnarray}  \label{Tmunu*}
T = T_{\mu}^{\mu}=\rho (r)c^2 -3p(r), p(r)=U_0\rho ^2(r),
\end{eqnarray}
and the field equations describing cylindrically symmetric string-type
solutions in general relativity,
\begin{equation}
R_{\mu \nu}=\frac{8\pi G}{c^4}\left(T_{\mu \nu}-\frac{1}{2}Tg_{\mu
\nu}\right),
\end{equation}
can be written as
\begin{eqnarray}  \label{eq1s}
\frac{(LKN^{\prime })^{\prime }}{NLK}= \frac{4\pi G}{c^4}\left(\rho
c^2+3p\right),
\end{eqnarray}%
\begin{eqnarray}  \label{eq2s}
\frac{N^{\prime \prime }}{N}+\frac{L^{\prime \prime }}{L}+\frac{K^{\prime
\prime }}{K}=\frac{4\pi G}{c^4}\left(p-\rho c^2\right),
\end{eqnarray}%
\begin{eqnarray}  \label{eq3s}
\frac{(NKL^{\prime })^{\prime }}{NLK}=\frac{4\pi G}{c^4}\left(p-\rho
c^2\right),
\end{eqnarray}%
\begin{eqnarray}  \label{eq4s}
\frac{(NLK^{\prime })^{\prime }}{NLK}=\frac{4\pi G}{c^4}\left(p-\rho
c^2\right).
\end{eqnarray}

Generally, the regularity of the geometry on the symmetry axis is imposed
via the initial conditions,
\begin{equation}
L(0)=0,L^{\prime }(0)=1,N(0)=1,N^{\prime }(0)=0.  \label{initcond}
\end{equation}

Next, we consider the relations that follow from the conservation of the
energy-momentum tensor, given by Eq.~(\ref{Tmunu}). Since all the components
of the energy-momentum tensor are independent of the coordinates $t$, $\phi$
and $z$, the relations
\begin{equation}
\nabla _{\mu }T_{t}^{\mu }=0,\nabla _{\mu }T_{\phi }^{\mu }=0,\nabla _{\mu
}T_{z}^{\mu }=0  \label{divzero}
\end{equation}
hold automatically. Furthermore, the divergence of $T_{\nu }^{\mu }$ can
be obtained in a general form as \cite{LaLi}
\begin{equation}
\nabla _{\mu }T_{\nu }^{\mu }=\frac{1}{\sqrt{-g}}\frac{\partial }{\partial
x^{\mu }}\left( \sqrt{-g}T_{\nu }^{\mu }\right) -\frac{1}{2}\frac{\partial
g_{\alpha \beta }}{\partial x^{\nu }}T^{\alpha \beta }.  \label{divem}
\end{equation}

It follows that Eqs.~(\ref{divzero}) are identically satisfied, with the
only potentially nonzero component of the divergence of the energy-momentum
tensor being given by
\begin{equation}
\nabla _{\mu }T_{r}^{\mu }=\frac{1}{\sqrt{-g}}\frac{d}{dr}\left( \sqrt{-g}%
T_{r}^{r}\right) -\frac{1}{2}\frac{\partial g_{\alpha \beta }}{\partial r}%
T^{\alpha \beta }=0,  \label{divem1}
\end{equation}
where $\sqrt{-g} = NKL$ and $\alpha = \beta \in \left\{r,\phi,z\right\}$ for
our choice of metric in Eq. (\ref{lineelement}), and energy-momentum tensor,
Eq. (\ref{Tmunu}).

Substituting Eq. (\ref{Tmunu}) into Eq. (\ref{divem1}), the
conservation equation for the BEC string takes the form
\begin{eqnarray}  \label{divemf}
\frac{dp}{dr} + \left(\rho c^2+p\right) \frac{N^{\prime }}{N}=0.
\end{eqnarray}

An upper bound for the density of the string arises from imposing the trace
energy condition $T=\rho c^2 - 3p\geq 0$, which restricts the range of
allowed densities to
\begin{equation} \label{TraceEC}
\rho \leq \frac{c^2}{3U_0}.
\end{equation}

To describe the physical characteristics of the BEC
string, we introduce the Tolman mass per unit length within the radius $r$
(hereafter, we will often use the phrase Tolman mass to refer to the
Tolman mass per unit length), $M(r)$, defined as
\begin{eqnarray}
&&M(r)=\frac{1}{c^2}\int{\left(\rho c^2-3U_0\rho ^2\right)\sqrt{-g}d^2x}=
\nonumber \\
&&2\pi\int_{r'=0}^{r}{\left[\rho \left(r'\right)-3\frac{U_0}{c^2}\rho ^2\left(r'\right)\right]N\left(r'\right)L\left(r'\right)K\left(r'\right) dr'}.\nonumber\\
\end{eqnarray}
The total Tolman mass of the string is defined as $M=\lim_{r\rightarrow
R_s}M\left(r\right)$, where $R_s$ is the radius of the string, which defines the vacuum boundary. We also introduce the parameter $W(r)$, which can
be related to the physical deficit angle in the space-time of the string, and
which is defined as \cite{Verbin}
\begin{eqnarray}
&&W(r) = -\frac{2\pi}{c^2} \int \left(\rho c^2 - p\right) \sqrt{-g}
d^2x=  \nonumber \\
&&-2\pi \int_{r'=0}^{r} \rho\left(r'\right) \left[1 - \frac{U_0}{c^2}\rho\left(r'\right)\right] N\left(r'\right)L\left(r'\right)K\left(r'\right) dr'.\nonumber\\
\end{eqnarray}
On the vacuum boundary of the string the function $W(r)$ has the finite value $W=\lim_{r\rightarrow
R_s}W(r)$. By assuming that the string {\it can extend to infinity}, so that $R_s\rightarrow \infty$, and that {\it the asymptotic form} of the metric for the Bose-Einstein Condensate  string is {\it flat}, from the field equations, Eqs.~(\ref{eq1s})-(\ref{eq4s}),
together with the definitions for $M$ and $W$, it follows that
\begin{eqnarray}
\left[N^{\prime }LK\right]_{r=0}^{\infty} &=& 2GM, \\
\left[NL^{\prime }K\right]_{r=0}^{\infty} &=& 2GW, \\
\left[NLK^{\prime }\right]_{r=0}^{\infty}&=& 2GW,
\end{eqnarray}
and
\begin{eqnarray}
\frac{L^{\prime }}{L} \bigg|_{r=0}^{\infty} = \frac{K^{\prime }}{K}\bigg|%
_{r=0}^{\infty}.
\end{eqnarray}
We then have
\begin{eqnarray}  \label{L'1}
L^{\prime }(\infty) = \frac{2GW + K(0)}{N(\infty)K(\infty)},
\end{eqnarray}
for $L^{\prime }(0) = N(0) = 1$, and the angular deficit in the cylindrical
geometry, due to the presence of the string, is given by \cite{Verbin}
\begin{eqnarray}  \label{Ang_Def1}
\Delta \phi = 2\pi(1-L^{\prime }(\infty)),
\end{eqnarray}
which can be numerically estimated by substituting values of $N(r)$ and $%
K(r) $, for large $r$. However, in order to do this for strings of finite width, we would require a
precise knowledge of the vacuum solution surrounding the string core. The
general form of this solution is well known: it is the Kasner solution, which
is the unique, cylindrically symmetric vacuum solution in general relativity \cite%
{Kasner,Harvey,exact-sol} but, for the sake brevity, we treat only the
interior solution for the BEC string in this paper. In order to estimate the
angular deficit using the formula given in Eq. (\ref{Ang_Def1}), we would
therefore need to match the Kasner space-time onto the interior solution at
the string boundary, $r=R_s$ (defined as the point beyond which the energy
and pressure density vanish, or become negative, so that $\rho c^2(R_s) =
p(R_s) = 0$), thereby fixing the numerical values of the Kasner parameters,
before taking the limit $r \rightarrow \infty$ to determine $L^{\prime
}(\infty)$. As a first approximation, however, we may substitute $L^{\prime
}(R_s)$ into Eq. (\ref{Ang_Def1}), in place of $L^{\prime }(\infty)$, and
use the formula
\begin{eqnarray}  \label{Ang_Def2}
\Delta \phi \approx 2\pi(1-L^{\prime }(R_s)),
\end{eqnarray}
from which we can obtain an order of magnitude estimate of the deficit angle
in the exterior BEC string geometry, using the numerical solutions of the
field equations for the string interior.

For the sake of notational simplicity, we now introduce the variables
\begin{equation}
\sqrt{-g} = \Sigma =NLK, \ H_t=\frac{N^{\prime }}{N},H_{\phi}=\frac{%
L^{\prime }}{L}, \ H_z=\frac{K^{\prime }}{K},
\end{equation}
and
\begin{equation}
H=\frac{1}{3}\left(H_t+H_{\phi}+H_z\right)=\frac{1}{3}\frac{\Sigma^{\prime }%
}{\Sigma}.
\end{equation}
The field equations describing an interaction energy dominated BEC string
then take the form
\begin{equation}  \label{tr1}
\frac{1}{\Sigma}\frac{d}{dr}\left(\Sigma H_t\right)=\frac{4\pi G}{c^4}%
\left(\rho c^2+3U_0\rho ^2\right),
\end{equation}
\begin{equation}  \label{tr2}
3\frac{dH}{dr}+H_t^2+H_{\phi}^2+H_z^2=\frac{4\pi G}{c^4}\left(U_0\rho
^2-\rho c^2\right),
\end{equation}
\begin{equation}  \label{tr3}
\frac{1}{\Sigma}\frac{d}{dr}\left(\Sigma H_i\right)=\frac{4\pi G}{c^4}%
\left(U_0\rho ^2-\rho c^2\right),i=\phi, z.
\end{equation}
From Eqs.~(\ref{tr3}) we immediately obtain
\begin{equation}  \label{H_rel}
H_{\phi}=H_z+\frac{C}{\Sigma},
\end{equation}
where $C$ is an arbitrary constant of integration. By adding Eqs.~(\ref{tr1}%
) and (\ref{tr3}) we have
\begin{eqnarray}  \label{V0}
\frac{3}{\Sigma}\frac{d}{dr}(\Sigma H)&=&\frac{\Sigma^{\prime \prime }}{%
\Sigma}=\frac{d}{dr}\frac{\Sigma^{\prime }}{\Sigma}+\left(\frac{%
\Sigma^{\prime }}{\Sigma}\right)^2=  
\frac{4\pi G}{c^4}\left(5U_0\rho ^2-\rho c^2\right), \nonumber \\&&
\end{eqnarray}
and from the conservation equation Eq.~(\ref{divemf}) it follows that
\begin{equation}  \label{Ht}
H_t=-\frac{2\left(U_0/c^2\right)\rho ^{\prime }}{1+\left(U_0/c^2\right)\rho},
\end{equation}
giving
\begin{equation}
N(r)=\frac{N_0}{\left[1+\left(U_0/c^2\right)\rho (r)\right]^2},
\end{equation}
where $N_0$ is an arbitrary constant of integration. Combining Eqs.~(\ref%
{tr1}) and (\ref{Ht}) gives
\begin{equation}  \label{38}
\frac{\Sigma^{\prime }}{\Sigma}H_t+H^{\prime }_t=\frac{4\pi G}{c^4}%
\left(\rho c^2+3U_0\rho ^2\right),
\end{equation}
and, with the help of Eq.~(\ref{Ht}), Eq.~(\ref{38}) allows us to express $%
\Sigma^{\prime }/\Sigma$ as
\begin{eqnarray}  \label{V1}
\frac{\Sigma^{\prime }}{\Sigma}&=&-\frac{\rho ^{\prime \prime }}{\rho
^{\prime }}+\frac{\left(U_0/c^2\right)\rho ^{\prime }}{1+\left(U_0/c^2%
\right)\rho }-  \nonumber \\
&&\frac{4\pi G}{c^2}\frac{\rho \left[1+\left(U_0/c^2\right)\rho \right]\left[%
1+\left(3U_0/c^2\right)\rho\right]}{2\left(U_0/c^2\right)\rho ^{\prime }}.
\end{eqnarray}

At this point we rescale the energy density, by introducing a new
dimensionless variable $\theta $ 
\footnote{This is not to be confused with the previous variable, which defined the form of the cylindrical Lane-Emden equation.}, so that
\begin{equation}
\theta (r)=\frac{U_0}{c^2}\rho (r),
\end{equation}
and, from here on, denote
\begin{equation}
\lambda =\frac{4\pi G}{U_0}.
\end{equation}
In the new variable $\theta $ the pressure distribution inside the string is
obtained as
\begin{equation}
p(r)=\frac{c^4}{U_0}\theta ^2(r),
\end{equation}
while the Tolman mass within a radius $r$ is given by
\begin{equation}
M(r)=\frac{2\pi c^2}{U_0}\int_{r'=0}^{r} {\theta \left(r '\right)\left[1-3\theta \left(r '\right)\right]\Sigma \left(r '\right)dr'},
\end{equation}
and the total Tolman mass of the string is $M=M\left(R_s\right)$.

The parameter $W(r)$ is given by
\begin{equation}
W(r)=-\frac{2\pi c^2}{U_0}\int_{r'=0}^{r} {\theta \left(r '\right)\left[1-\theta \left(r '\right)\right]\Sigma \left(r '\right)dr'},
\end{equation}
and the surface value of $W(r)$ is $W=W\left(R_s\right)$. Then Eq.~(\ref{V1})
can be written as
\begin{equation}  \label{V1*}
\frac{\Sigma^{\prime }}{\Sigma}=-\frac{\theta ^{\prime \prime }}{\theta
^{\prime }}+\frac{\theta ^{\prime }}{1+\theta }- \lambda \frac{\theta
\left(1+\theta \right)\left(1+3\theta \right)}{2\theta ^{\prime }}.
\end{equation}
Thus, by combining Eq.~(\ref{V0}), written as
\begin{equation}
\frac{d}{dr}\left(\frac{\Sigma^{\prime }}{\Sigma}\right)+\left(\frac{%
\Sigma^{\prime }}{\Sigma}\right)^2=\lambda \theta (5\theta -1),
\end{equation}
with Eq. (\ref{V1*}), we obtain the following third-order ordinary nonlinear
differential equation for the energy density distribution inside the BEC
string,
\begin{eqnarray}  \label{eqrho}
&&\frac{\theta ^{\prime \prime \prime }}{\theta ^{\prime }}-\frac{2\theta
^{\prime \prime 2}}{\theta ^{\prime 2}}- \left[ \frac{3\lambda \theta
(\theta +1)(3\theta +1)}{2\theta ^{\prime 2}}-\frac{1}{\theta +1}\right]
\theta ^{\prime \prime }-  \nonumber \\
&&\frac{\lambda ^{2}\theta ^{2}(\theta +1)^{2}(3\theta +1)^{2}}{4\theta
^{\prime 2}}+\frac{1}{2}\lambda \left( 25\theta ^{2}+8\theta +1\right) =0.
\nonumber \\
\end{eqnarray}
However, instead of studying Eq.~(\ref{eqrho}) numerically, it is more
advantageous to study the following equivalent autonomous system of
differential equations,
\begin{eqnarray}
\Sigma^{\prime }&=&u,  \label{40} \\
u^{\prime }&=&\lambda \theta (5\theta -1)\Sigma, \\
\theta ^{\prime }&=&v, \\
v^{\prime }&=&-\frac{u}{\Sigma}v+\frac{v^2}{1+\theta }-\lambda \frac{\theta
\left(1+\theta \right)\left(1+3\theta\right)}{2}, \\
K^{\prime }&=&k, \\
k^{\prime }&=&\frac{k^2}{K}-\frac{u}{\Sigma}k+\lambda \theta (\theta -1)K.
\label{44}
\end{eqnarray}
where the expression for $k^{\prime }$ follows directly from Eq.~(\ref{tr3}).

The system of Eqs.~(\ref{40})--(\ref{44}) must be considered with the
initial conditions $\Sigma(0)=\Sigma_0$, $u(0)=\Sigma^{\prime
}(0)=3H(0)\Sigma(0)$, $\theta (0)=\left(U_0/c^2\right)\rho _0$, $K(0)=K_0$, $%
v(0)=\theta ^{\prime }(0)=-\left[\left(1+\theta _0\right)/2\right]%
\left.\left(N^{\prime }/N\right)\right|_{r=0}$ and $k(0)=K^{\prime
}(0)=k_0^{\prime }$, respectively. Once the solution of the system given in
Eqs. (\ref{40})--(\ref{44}) is obtained, the variation of metric tensor
component $L$ can be obtained from the equation $L=\Sigma/NK$. The initial
condition $N(0)=1$, imposed on the metric function $N$, determines the
integration constant $N_0$ as $N_0=\left[1+\left(U_0/c^2\right)\rho _0\right]%
^2$.

\begin{figure}[h]
\caption{Variation of $\Sigma(r) = \sqrt{-g}$, for the BEC string
space-time, for different values of $U_0$: $U_0=10^6\;\mathrm{cm^5/g\;s^2}$
(solid curve), $U_0=10^{6.04}\;\mathrm{cm^5/g\;s^2}$ (dotted curve), $%
U_0=10^{6.08}\;\mathrm{cm^5/g\;s^2}$ (short dashed curve), $U_0=10^{6.12}\;%
\mathrm{cm^5/g\;s^2}$ (dashed curve), and $U_0=10^{6.16}\;\mathrm{cm^5/g\;s^2%
}$ (long dashed curve), respectively.}
\label{fig1}\centering
\includegraphics[width=8cm]{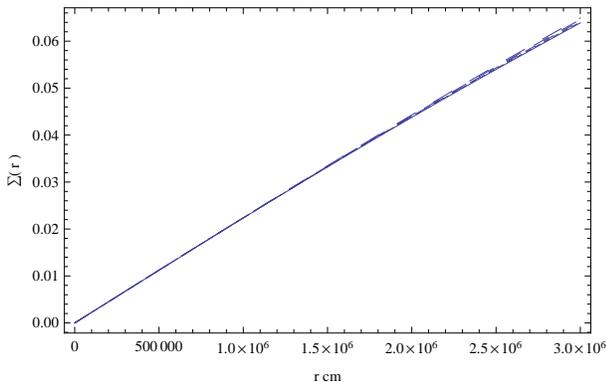}
\end{figure}

\begin{figure}[h]
\caption{Variation of the dimensionless density $\theta(r) = U_0%
\rho(r)/c^2$ of the BEC string, for different values of $U_0$: $%
U_0=10^6\;\mathrm{cm^5/g\;s^2}$ (solid curve), $U_0=10^{6.04}\;\mathrm{%
cm^5/g\;s^2}$ (dotted curve), $U_0=10^{6.08}\;\mathrm{cm^5/g\;s^2}$ (short
dashed curve), $U_0=10^{6.12}\;\mathrm{cm^5/g\;s^2}$ (dashed curve), and $%
U_0=10^{6.16}\;\mathrm{cm^5/g\;s^2}$ (long dashed curve), respectively. }
\label{fig2}\centering
\includegraphics[width=8cm]{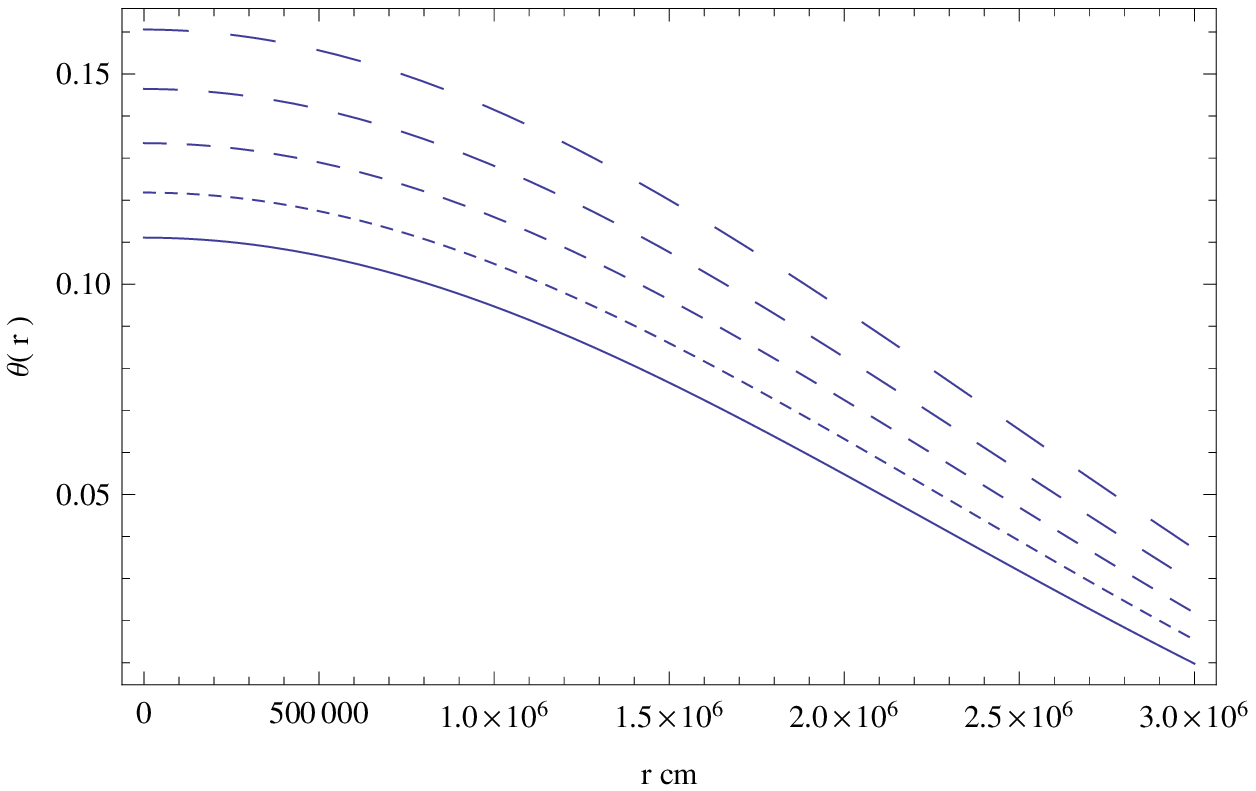}
\end{figure}

\begin{figure}[h]
\caption{Variation of the dimensionless pressure $\theta%
^2(r)=\left(U_0/c^4\right)p(r)$ of the BEC string, for different values of $%
U_0$: $U_0=10^6\;\mathrm{cm^5/g\;s^2}$ (solid curve), $U_0=10^{6.04}\;%
\mathrm{cm^5/g\;s^2}$ (dotted curve), $U_0=10^{6.08}\;\mathrm{cm^5/g\;s^2}$
(short dashed curve), $U_0=10^{6.12}\;\mathrm{cm^5/g\;s^2}$ (dashed curve),
and $U_0=10^{6.16}\;\mathrm{cm^5/g\;s^2}$ (long dashed curve), respectively.
}
\label{fig3}\centering
\includegraphics[width=8cm]{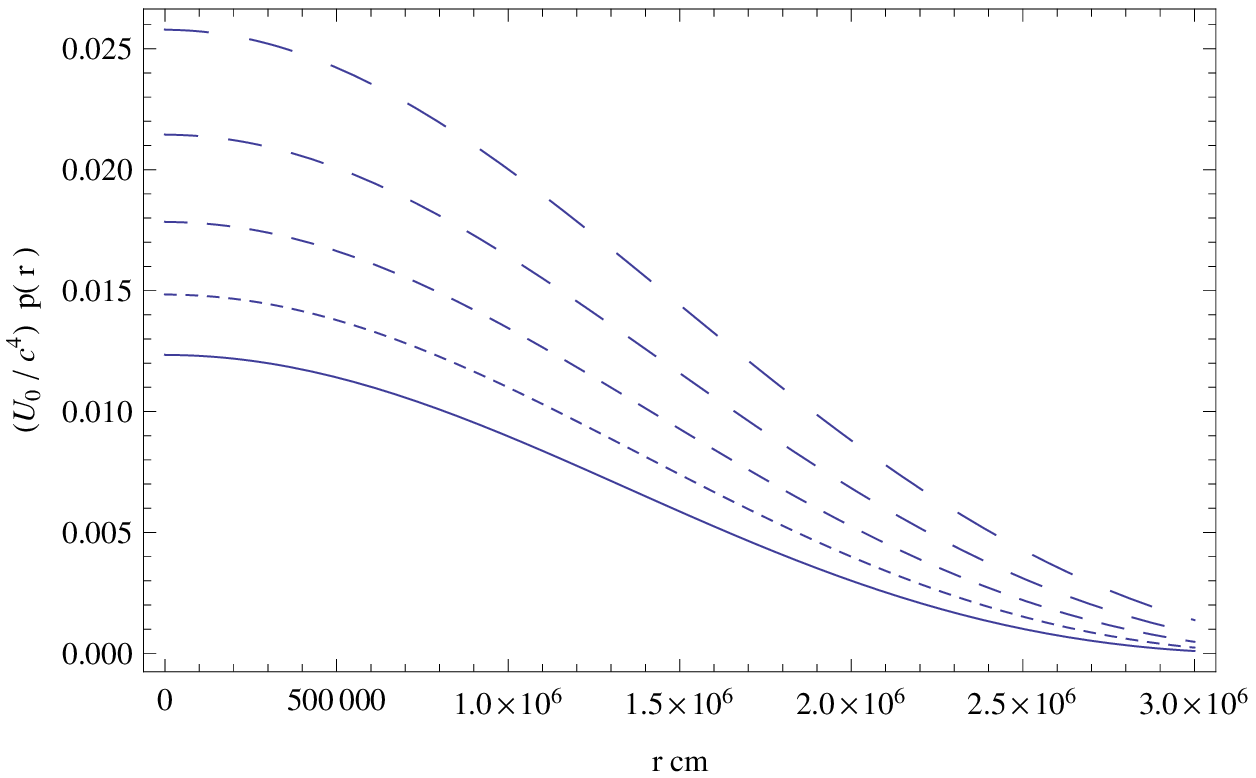}
\end{figure}

\begin{figure}[h]
\caption{Variation of the function $N(r)/N_0 \propto \sqrt{g_{tt}}$,
for the BEC string space-time, for different values of $U_0$: $U_0=10^6\;%
\mathrm{cm^5/g\;s^2}$ (solid curve), $U_0=10^{6.04}\;\mathrm{cm^5/g\;s^2}$
(dotted curve), $U_0=10^{6.08}\;\mathrm{cm^5/g\;s^2}$ (short dashed curve), $%
U_0=10^{6.12}\;\mathrm{cm^5/g\;s^2}$ (dashed curve), and $U_0=10^{6.16}\;%
\mathrm{cm^5/g\;s^2}$ (long dashed curve), respectively. }
\label{fig4}\centering
\includegraphics[width=8cm]{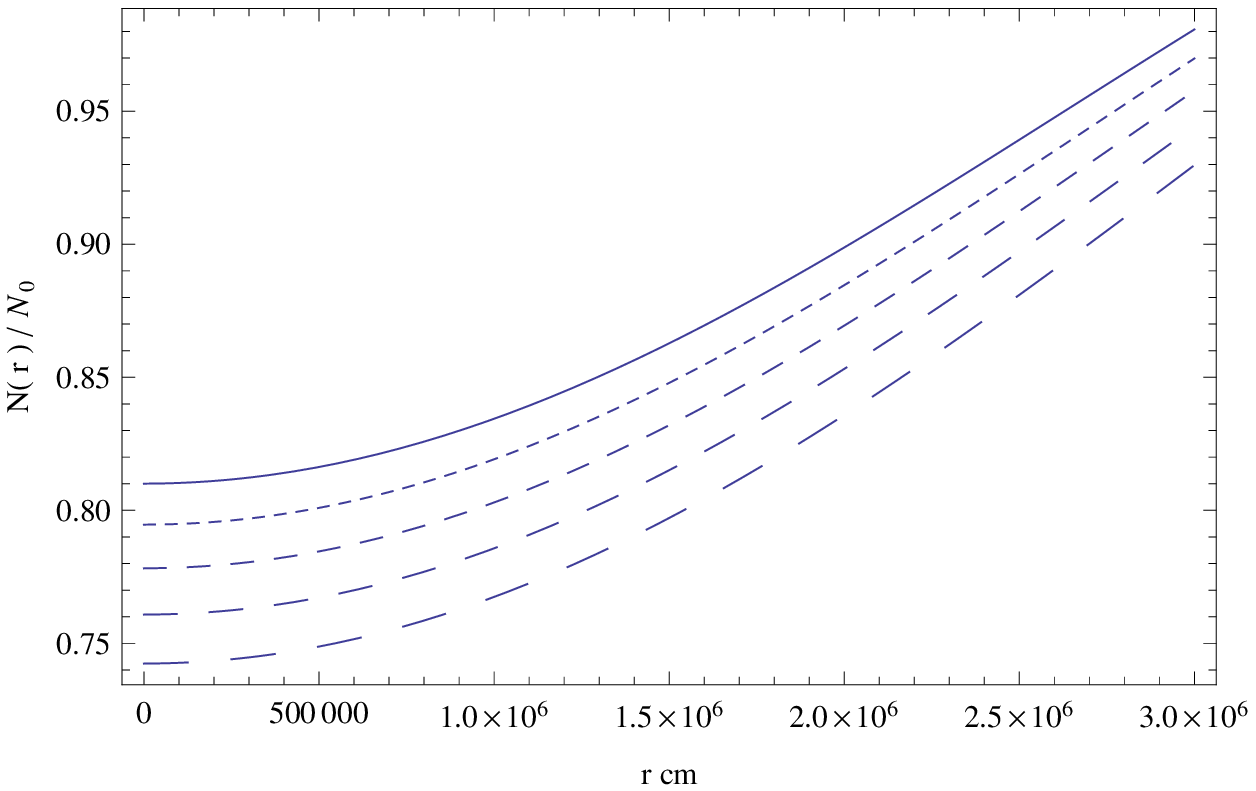}
\end{figure}

\begin{figure}[h]
\caption{Variation of $L(r) = \sqrt{-g_{\phi \phi}}$%
, for the BEC string space-time, for different values of $U_0$: $U_0=10^6\;%
\mathrm{cm^5/g\;s^2}$ (solid curve), $U_0=10^{6.04}\;\mathrm{cm^5/g\;s^2}$
(dotted curve), $U_0=10^{6.08}\;\mathrm{cm^5/g\;s^2}$ (short dashed curve), $%
U_0=10^{6.12}\;\mathrm{cm^5/g\;s^2}$ (dashed curve), and $U_0=10^{6.16}\;%
\mathrm{cm^5/g\;s^2}$ (long dashed curve), respectively. }
\label{fig5}\centering
\includegraphics[width=8cm]{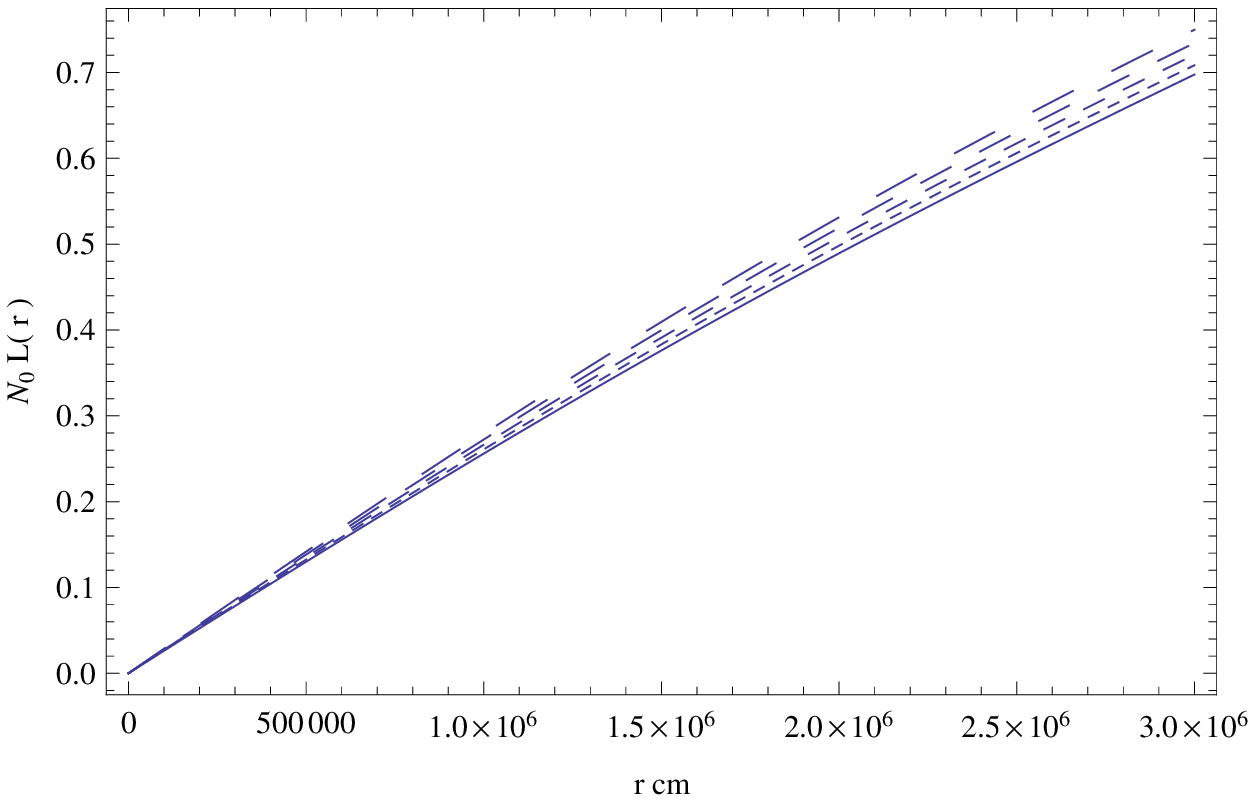}
\end{figure}

\begin{figure}[h]
\caption{Variation of $K(r) = \sqrt{-g_{zz}}$, for the BEC string
space-time, for different values of $U_0$: $U_0=10^6\;\mathrm{cm^5/g\;s^2}$
(solid curve), $U_0=10^{6.04}\;\mathrm{cm^5/g\;s^2}$ (dotted curve), $%
U_0=10^{6.08}\;\mathrm{cm^5/g\;s^2}$ (short dashed curve), $U_0=10^{6.12}\;%
\mathrm{cm^5/g\;s^2}$ (dashed curve), and $U_0=10^{6.16}\;\mathrm{cm^5/g\;s^2%
}$ (long dashed curve), respectively. }
\label{fig6}\centering
\includegraphics[width=8cm]{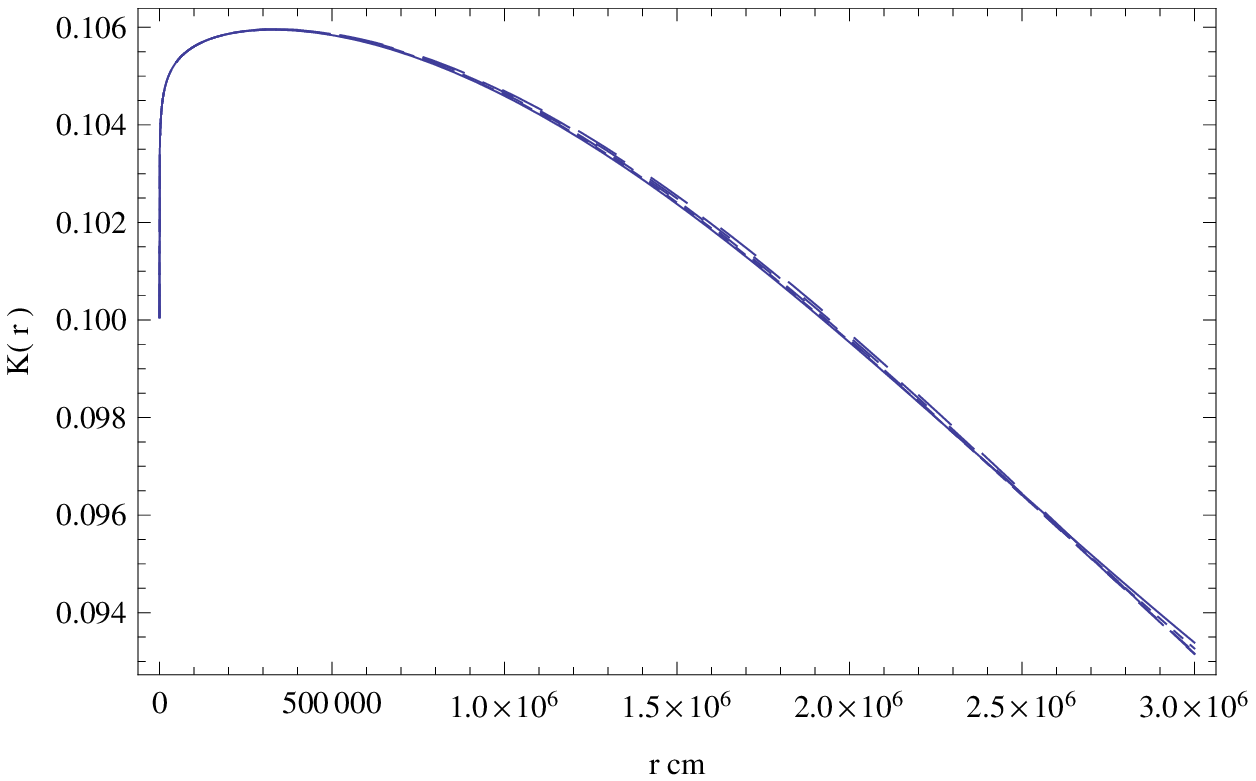}
\end{figure}

\begin{figure}[h]
\caption{Variation of the Tolman mass $M(r)$ of the BEC string, for
different values of $U_0$: $U_0=10^6\;\mathrm{cm^5/g\;s^2}$ (solid curve), $%
U_0=10^{6.04}\;\mathrm{cm^5/g\;s^2}$ (dotted curve), $U_0=10^{6.08}\;\mathrm{%
cm^5/g\;s^2}$ (short dashed curve), $U_0=10^{6.12}\;\mathrm{cm^5/g\;s^2}$
(dashed curve), and $U_0=10^{6.16}\;\mathrm{cm^5/g\;s^2}$ (long dashed
curve), respectively. }
\label{fig7}\centering
\includegraphics[width=8cm]{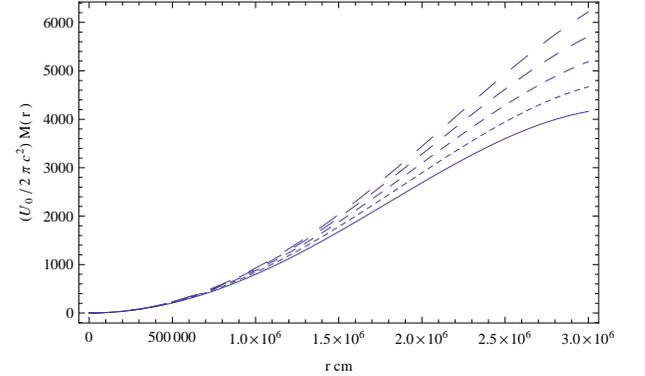}
\end{figure}

\begin{figure}[h]
\caption{Variation of the angular deficit parameter $W(r)$ of the BEC
string, for different values of $U_0$: $U_0=10^6\;\mathrm{cm^5/g\;s^2}$
(solid curve), $U_0=10^{6.04}\;\mathrm{cm^5/g\;s^2}$ (dotted curve), $%
U_0=10^{6.08}\;\mathrm{cm^5/g\;s^2}$ (short dashed curve), $U_0=10^{6.12}\;%
\mathrm{cm^5/g\;s^2}$ (dashed curve), and $U_0=10^{6.16}\;\mathrm{cm^5/g\;s^2%
}$ (long dashed curve), respectively. }
\label{fig71}\centering
\includegraphics[width=8cm]{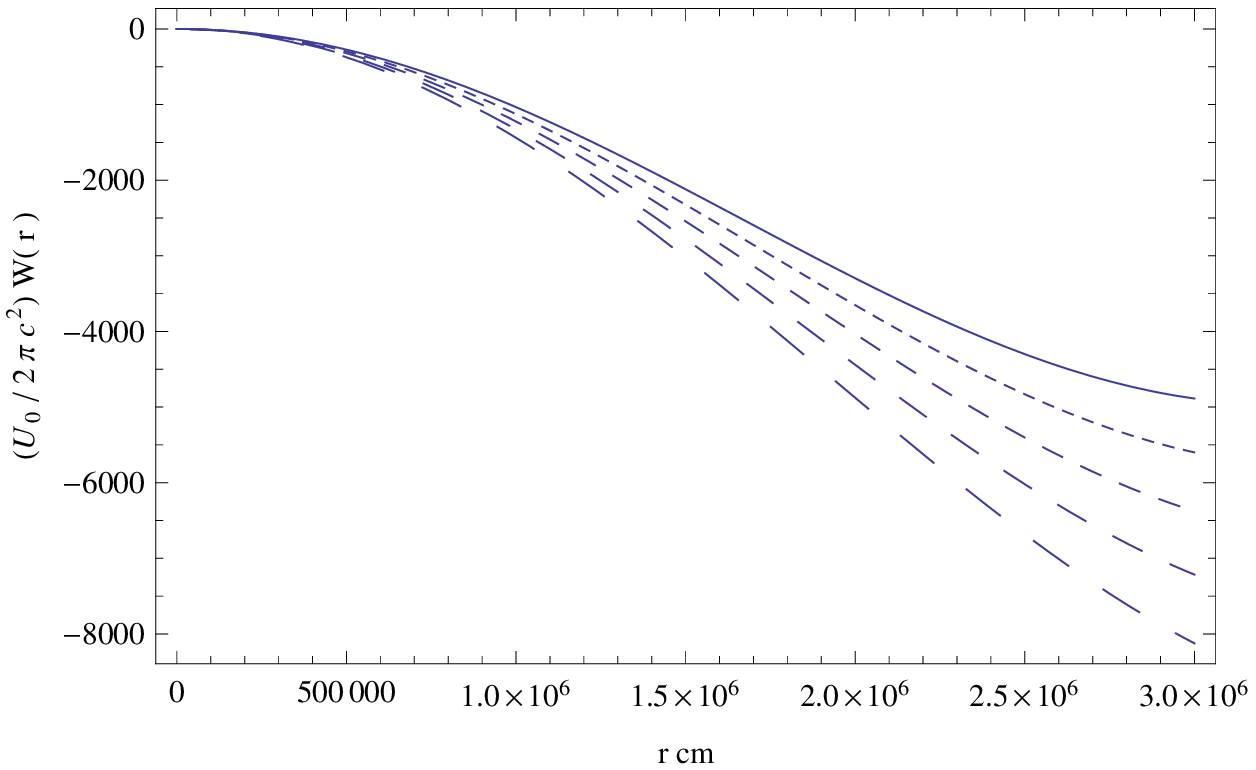}
\end{figure}

The variation of the functions $\Sigma(r) = \sqrt{-g}$, $\theta(r) \propto
\rho(r)$, $\theta^2(r) \propto p(r)$, $N(r) = \sqrt{g_{tt}}$, $L(r) = \sqrt{%
-g_{\phi\phi}}$ and $K(r) = \sqrt{-g_{zz}}$, obtained from the numerical
solution of Eqs. (\ref{40})-(\ref{44}), together with appropriate (example)
numerical values for the initial conditions, are represented in Figs.~\ref%
{fig1} - \ref{fig6}, respectively. The variation of the Tolman mass $M(r)$ is given
in Fig.~\ref{fig7}, while the variation of the parameter $W(r)$ is presented in
Fig.~\ref{fig71}. In Figs. ~\ref{fig1} - \ref{fig71} the initial
conditions used to numerically integrate the system, Eqs.~(\ref{40})-(\ref%
{44}), were $\Sigma(0)=10^{-8}$, $\rho (0)=10^{14}$ g/cm$^2$, $H(0)=0.75$ cm%
$^{-1}$, $\left.N^{\prime }/N\right|_{r=0}=10^{-5}$ cm$^{-1}$, $K(0)=0.1$, $%
K^{\prime }(0)=0.001$, $M(0)=0$ and $W(0)=0$, respectively.

As one can see from Fig.~\ref{fig1}, for small $r$, the function $\Sigma(r) = \sqrt{-g}$
appears to be (approximately) proportional to $r$, but with constant of
proportionality much less than one (compared to $\Sigma(r) = \sqrt{-g} = r$
for a flat conical geometry, such as that obtained for a vacuum
string). The dimensionless density $\theta(r) = U_0\rho(r)/c^2$, presented
in Fig.~\ref{fig2}, appears, roughly, to be a Bell-shaped curve. It
monotonically decreases from a maximum central value at $r=0$ (as expected
intuitively), and reaches the value $\theta =0$ at some finite value of $r$,
which defines the radius of the string $R_s$, $\theta \left(R_s\right)=0$.
In the examples considered, the radius of the string is of the order of $%
R_s\approx 3\times 10^6\;\mathrm{cm}\approx 30\;\mathrm{km}$. The physical
pressure, $p(r) \propto \theta^2(r)$, of the Bose-Einstein condensate that
forms the string interior, which is shown in Fig.~\ref{fig3}, also becomes
zero for $r=R_s$, that is, at the vacuum boundary of the string. The
behavior of the functions that determine the metric tensor components,
depicted in Figs.~\ref{fig4}-\ref{fig6}, show very different variation with
respect to the radial coordinate: $N(r)/N_0 \propto \sqrt{g_{tt}}$ is a
monotonically increasing function of $r$, and is roughly proportional to $r$
at large radii. For small $r$, $L(r) = \sqrt{-g_{\phi \phi}}$ is also roughly
proportional to $r$, but with a constant of proportionality much less than
one. For large $r$, $L(r)$ varies according to some power of $r$, $r^{\delta}$, where $0 < \delta < 1$.
This is in sharp contrast with the
flat conical geometry of a vacuum string, for which $L(r) = \sqrt{-g_{\phi
\phi}} = r$. The metric function $K(r) = \sqrt{-g_{zz}}$ rises sharply from
its initial value, and peaks rapidly, before decreasing monotonically.
This behavior is again very different, as compared to $K(r) =
\sqrt{-g_{zz}} = 1$, which corresponds to flat conical Minkowski space. The
Tolman mass function, presented in Fig.~\ref{fig7}, is monotonically
increasing, and reaches its maximum value at $r=R_s$, giving a total mass per
unit length for the BEC string (in the examples considered) of the order $M\approx 4-6\times 10^3\times
\left(2\pi c^2\right)/U_0\approx 2-3\times 10^{19}$ g. This is extremely
small (for reasonable values of $U_0$), compared to the masses of
possible BEC matter neutron stars, for example, which are of the order of $%
M\approx 2.8\times 10^{33}$ g. The angular deficit parameter $W(r)$, plotted in
Fig.~\ref{fig71}, has negative values inside the string, and
monotonically decreases as it approaches the string boundary.

This latter observation is perhaps the most intriguing aspect of the BEC
string as it allows, in principle, for an ``angle excess" in the resulting
space-time, according to Eqs. (\ref{L'1})-(\ref{Ang_Def2}). The term angle
excess is used to describe conical geometries in which the deficit angle
exceeds $2\pi$. Though these may be considered unphysical, Visser has
suggested that, instead, such configurations correspond to negative mass
strings (composed of exotic matter), which may be capable of supporting a
traversable wormhole \cite{Visser1989}. An example of such an exotic
string, originally proposed in \cite{WormholeStrings1}, is one with
vanishing radial and azimuthal stresses ($T_r = T_{\phi} = 0, \ \forall r$),
for which the mass per unit length $\mu$ is equal to the longitudinal
tension $T_z$, both of which are negative, i.e. $\mu = T_z < 0$, and further
work has since been carried out in relation to this idea \cite%
{WormholeStrings2}. The exoticness of such an object may be judged against
the corresponding condition for Nambu-Goto or vacuum string, $\mu = -T_z > 0$%
. Since the BEC string, considered here, clearly has positive mass and
positive pressure (i.e. negative tension) in the longitudinal direction,
this raises the intriguing question as to whether exotic negative mass
objects are really required to support traversable wormholes, or whether
other cylindrically symmetric mass distributions, obeying reasonable sets of
energy conditions (for example, the trace energy condition Eq. (\ref{TraceEC})), may be able to behave exotically, given the right
initial conditions.


\subsection{Quantum pressure dominated Bose-Einstein condensate strings}

\label{Sect.IIID}

In this section we assume that the BEC string is supported by its quantum
pressure $p_Q$, given by Eq.~(\ref{quantpress}), which satisfies the
condition $p_Q>>p=U_0\rho ^2$. By adopting the Newtonian approximation for
the quantum regime, and assuming that the gravitational field does not
affect the fundamental quantum properties of the system, as formulated in
the standard Hilbert space approach, the three-dimensional Laplacian
operator, $\vec{\nabla}^2$, for cylindrically symmetry systems, is given by
\begin{equation}
\Delta =\frac{1}{r}\frac{d}{dr}\left(r\frac{d}{dr}\right).
\end{equation}

Therefore, the gravitational field equations describing the quantum pressure
supported BEC string take the form
\begin{equation}  \label{q1}
\frac{1}{\Sigma }\frac{d}{dr}\left(\Sigma H_t\right)= \frac{4\pi G}{c^2}\rho %
\left[1-3\frac{\hbar ^2}{4m^2c^2}\frac{1}{r}\frac{d}{dr}\left(r\frac{d}{dr}%
\ln \frac{\rho}{\rho _c} \right)\right],
\end{equation}
\begin{eqnarray}  \label{q2}
&&3\frac{dH}{dr}+H_t^2+H_{\phi}^2+H_z^2=  \nonumber \\
&&-\frac{4\pi G}{c^2}\rho \left[1+\frac{\hbar ^2}{4m^2c^2}\frac{1}{r}\frac{d%
}{dr}\left(r\frac{d}{dr}\ln \frac{\rho }{\rho _c} \right)\right],
\end{eqnarray}
\begin{eqnarray}  \label{q3}
\frac{1}{\Sigma }\frac{d}{dr}\left(\Sigma H_i\right)&=&-\frac{4\pi G}{c^2}%
\rho \left[1+\frac{\hbar ^2}{4m^2c^2}\frac{1}{r}\frac{d}{dr}\left(r\frac{d}{%
dr}\ln \frac{\rho}{\rho _c} \right)\right],  \nonumber \\
&&i=\phi, z,
\end{eqnarray}
where $\rho _c$ is the central density of the string. The corresponding
Tolman mass is given by
\begin{equation}
M(r)=2\pi \int_0^{r}{\rho \left[1+3\frac{\hbar }{4m^2c^2}\frac{1}{r}\frac{d}{%
dr}\left(r\frac{d}{dr}\ln \frac{\rho }{\rho _c}\right)\right]\Sigma dr}.
\end{equation}
As in the case of the interaction energy dominated BEC string, the relation
between $H_{\phi}$ and $H_z$ is again given by Eq. (\ref{H_rel}). %
By introducing the dimensionless variables $\zeta$ and $\tau$, defined as
\begin{equation}
r=\frac{\hbar}{2mc}\zeta, \ \tau =\frac{\rho }{\rho _c},
\end{equation}
and by denoting
\begin{equation}
\lambda _Q=\frac{\pi G\hbar ^2}{m^2c^4}\rho _c,
\end{equation}
the field equations describing the quantum pressure dominated BEC string
take the following dimensionless form:
\begin{equation}  \label{dq1}
\frac{1}{\Sigma }\frac{d}{d\zeta }\left(\Sigma {\mathcal{H}}%
_t\right)=\lambda _Q\tau \left[1-3\frac{1}{\zeta}\frac{d}{d\zeta }%
\left(\zeta \frac{d}{d\zeta }\ln \tau \right)\right],
\end{equation}
\begin{eqnarray}  \label{dq2}
3\frac{d{\mathcal{H}}}{d\zeta }+{\mathcal{H}}_t^2+{\mathcal{H}}_{\phi}^2+{%
\mathcal{H}}_z^2= -\lambda _Q\tau \left[1+\frac{1}{\zeta }\frac{d}{d\zeta }%
\left(\zeta \frac{d}{d\zeta }\ln \tau \right)\right],  \nonumber \\
\end{eqnarray}
\begin{equation}  \label{dq3}
\frac{1}{\Sigma }\frac{d}{d\zeta }\left(\Sigma {\mathcal{H}}%
_i\right)=-\lambda _Q\tau \left[1+\frac{1}{\zeta }\frac{d}{d\zeta }%
\left(\zeta \frac{d}{d\zeta }\ln \tau \right)\right],i=\phi, z,
\end{equation}
where the ${\mathcal{H}}_{\nu}$, $\nu \in \left\{t,\phi,z\right\}$ and ${%
\mathcal{H}}$ are now defined in terms of the derivatives with respect to $%
\zeta$, i.e. $\mathcal{H}_{\nu} = (\hbar/2mc)H_{\nu}$, so that $\mathcal{H}%
_t=(1/N)\left(dN/d\zeta \right)$ etc.

The energy conservation equation for the quantum pressure becomes
\begin{equation}
\frac{d}{d\zeta }\left[ \tau \frac{1}{\zeta }\frac{d}{d\zeta }\left( \zeta
\frac{d}{d\zeta }\ln \tau \right) \right] =\tau \left[ 1-\frac{1}{\zeta }%
\frac{d}{d\zeta }\left( \zeta \frac{d}{d\zeta }\ln \tau \right) \right] {%
\mathcal{H}}_{t},
\end{equation}%
and its dimensionless equivalent, $P_{Q}=p_{Q}/\rho _{c}c^{2} $, may be
written as 
\begin{equation}
P_{Q}=-\tau \frac{1}{\zeta }\frac{d}{d\zeta }\left( \zeta \frac{d}{d\zeta }%
\ln \tau \right) .
\end{equation}%
By adding Eqs.~(\ref{dq1}) and (\ref{dq3}) we obtain
\begin{equation}
\frac{1}{\Sigma }\frac{d^{2}\Sigma }{d\zeta ^{2}}=-\lambda _{Q}\tau \left[
1+5\frac{1}{\zeta }\frac{d}{d\zeta }\left( \zeta \frac{d}{d\zeta }\ln \tau
\right) \right] ,
\end{equation}%
while Eqs.~(\ref{dq1}) and (\ref{dq3}) for ${\mathcal{H}}_{z}$ can be
written as
\begin{equation}
\frac{d{\mathcal{H}}_{t}}{d\zeta }+\frac{1}{\Sigma }\frac{d\Sigma }{d\zeta }{%
\mathcal{H}}_{t}=\lambda _{Q}\tau \left[ 1-3\frac{1}{\zeta }\frac{d}{d\zeta }%
\left( \zeta \frac{d}{d\zeta }\ln \tau \right) \right] ,
\end{equation}%
and
\begin{equation}
\frac{d{\mathcal{H}}_{z}}{d\zeta }+\frac{1}{\Sigma }\frac{d\Sigma }{d\zeta }{%
\mathcal{H}}_{z}=-\lambda _{Q}\tau \left[ 1+\frac{1}{\zeta }\frac{d}{d\zeta }%
\left( \zeta \frac{d}{d\zeta }\ln \tau \right) \right] ,
\end{equation}%
respectively. Equivalently, the system of field equations obtained above,
which may be viewed as a relativistic generalization of the cylindrical
Lane-Emden equation discussed in Sect. \ref{Sect.IIB}, can then be
formulated as a system of first-order differential equations, i.e.
\begin{equation}
\frac{d\tau }{d\zeta }=a,\frac{d\Sigma }{d\zeta }=\sigma ,\frac{dN}{d\zeta }%
=n,\frac{dK}{d\zeta }=k,  \label{iq1}
\end{equation}%
\begin{equation}
\frac{da}{d\zeta }=\tau u+\frac{a^{2}}{\tau }-\frac{a}{\zeta },
\end{equation}%
\begin{equation}
\frac{du}{d\zeta }=(1-u)\frac{n}{N}-\frac{au}{\tau },  \label{iq2}
\end{equation}%
\begin{equation}
\frac{dv}{d\zeta }=-\lambda _{Q}\tau (1+5u)\Sigma ,  \label{iq3}
\end{equation}%
\begin{equation}
\frac{dn}{d\zeta }=\lambda _{Q}\tau (1-3u)N+\frac{n^{2}}{N}-\frac{\sigma }{%
\Sigma }n,  \label{iq3_1}
\end{equation}%
\begin{equation}
\frac{dk}{d\zeta }=-\lambda _{Q}\theta (1+u)K+\frac{k^{2}}{K}-\frac{\sigma }{%
\Sigma }{k},  \label{iq4}
\end{equation}%
which must be integrated subject to the initial conditions $\tau (0)=1$, $%
a(0)=\tau ^{\prime }(0)=\tau _{0}^{\prime }$, $u(u)=u_{0}^{\prime }$, $%
N(0)=1 $, $n(0)=N^{\prime }(0)=0$, $K(0)=K_{0}$, $k(0)=K^{\prime
}(0)=K_{0}^{\prime }$, $\Sigma (0)=\Sigma _{0}$ and $\sigma (0)=\Sigma
^{\prime }(0)=\sigma _{0}^{\prime }$, where a prime now indicates
differentiation with respect to $\zeta $. The dimensionless form of the
Tolman mass, $m(\zeta )=M(\zeta )/\left( \pi \hbar ^{2}\rho
_{c}/2m^{2}c^{2}\right) $, is given by
\bea
m(\zeta )&=&2\pi \int_{\zeta' =0}^{\zeta }\tau \left(\zeta '\right)\left[ 1+3\frac{1}{\zeta' }\frac{%
d}{d\zeta' }\left( \zeta' \frac{d}{d\zeta' }\ln \tau \left(\zeta '\right)\right) \right] \times \nonumber\\
&&\Sigma \left(\zeta '\right)
d\zeta' ,
\eea%
while the dimensionless angular deficit $w(\zeta )=W(\zeta )/\left( \pi
\hbar ^{2}\rho _{c}/2m^{2}c^{2}\right) $ is obtained as
\bea
w(\zeta )&=&-2\pi \int_{\zeta' =0}^{\zeta }\tau \left(\zeta '\right)\left[ 1+\frac{1}{\zeta' }\frac{%
d}{d\zeta' }\left( \zeta' \frac{d}{d\zeta' }\ln \tau \left(\zeta '\right)\right) \right] \times \nonumber\\
&&\Sigma \left(\zeta '\right)
d\zeta' .
\eea

\begin{figure}[h]
\caption{Variation of $\Sigma = \sqrt{-g}$, as a function of $%
\zeta$, in the space-time of the quantum pressure supported Bose
Einstein condensate string, for different values of $\lambda _Q$: $%
\lambda _Q=0.10$ (solid curve), $\lambda _Q=0.14$ (dotted
curve), $\lambda _Q=0.18$ (short dashed curve), $\lambda %
_Q=0.20 $ (dashed curve), and $\lambda _Q=0.22$ (long dashed curve),
respectively. }
\label{fig8}\centering
\includegraphics[width=8cm]{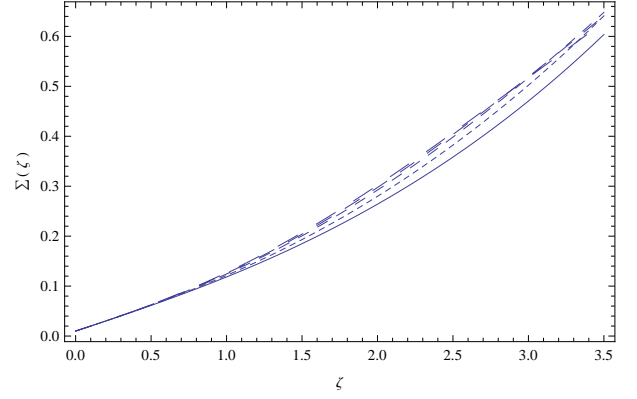}
\end{figure}

\begin{figure}[h]
\caption{Variation of the dimensionless density $\tau $, as a
function of $\zeta$, of the quantum pressure dominated Bose-Einstein
condensate string, for different values of $\lambda _Q$: $%
\lambda _Q=0.10$ (solid curve), $\lambda _Q=0.14$ (dotted curve), $%
\lambda _Q=0.18$ (short dashed curve), $\lambda _Q=0.20 $
(dashed curve), and $\lambda _Q=0.22$ (long dashed curve),
respectively. }
\label{fig9}\centering
\includegraphics[width=8cm]{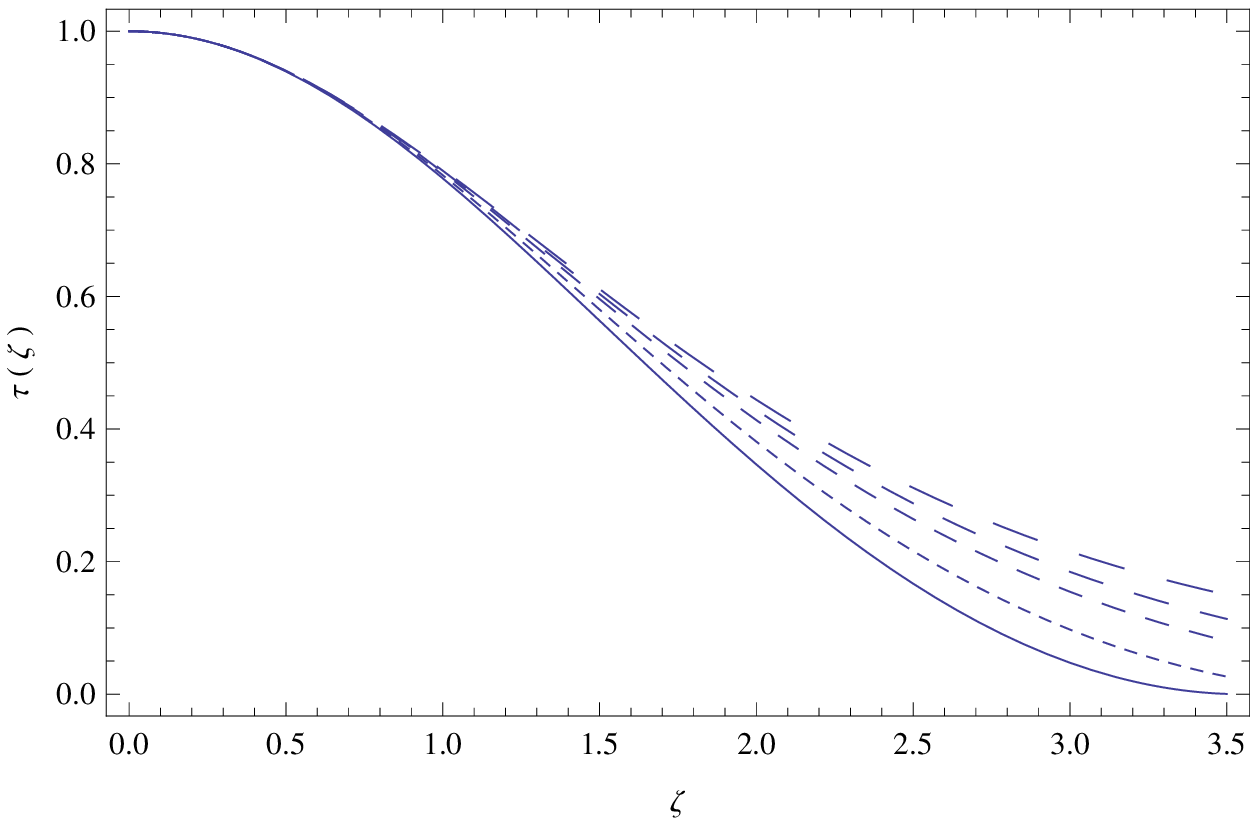}
\end{figure}

\begin{figure}[h]
\caption{Variation of the dimensionless quantum pressure $P_Q$, as a
function of $\zeta$, of the quantum pressure dominated Bose-Einstein
condensate string, for different values of $\lambda _Q$: $%
\lambda _Q=0.10$ (solid curve), $\lambda _Q=0.14$ (dotted curve), $%
\lambda _Q=0.18$ (short dashed curve), $\lambda _Q=0.20 $
(dashed curve), and $\lambda _Q=0.22$ (long dashed curve),
respectively. }
\label{fig10}\centering
\includegraphics[width=8cm]{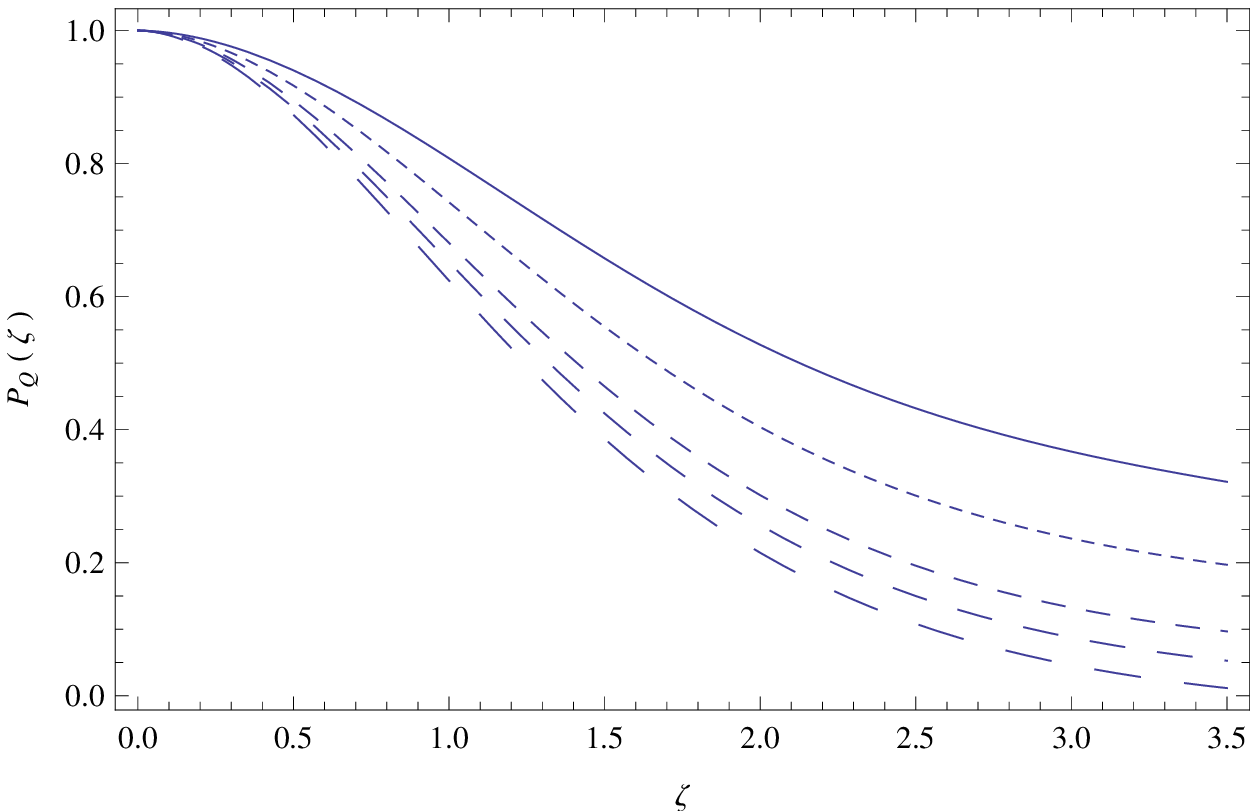}
\end{figure}

\begin{figure}[h]
\caption{Variation of the $N(\zeta) = \sqrt{g_{tt}}$ for the
space-time of the quantum pressure dominated Bose-Einstein condensate string,
for different values of $\lambda _Q$: $\lambda _Q=0.10$
(solid curve), $\lambda _Q=0.14$ (dotted curve), $\lambda %
_Q=0.18$ (short dashed curve), $\lambda _Q=0.20 $ (dashed curve),
and $\lambda _Q=0.22$ (long dashed curve), respectively. }
\label{fig11}\centering
\includegraphics[width=8cm]{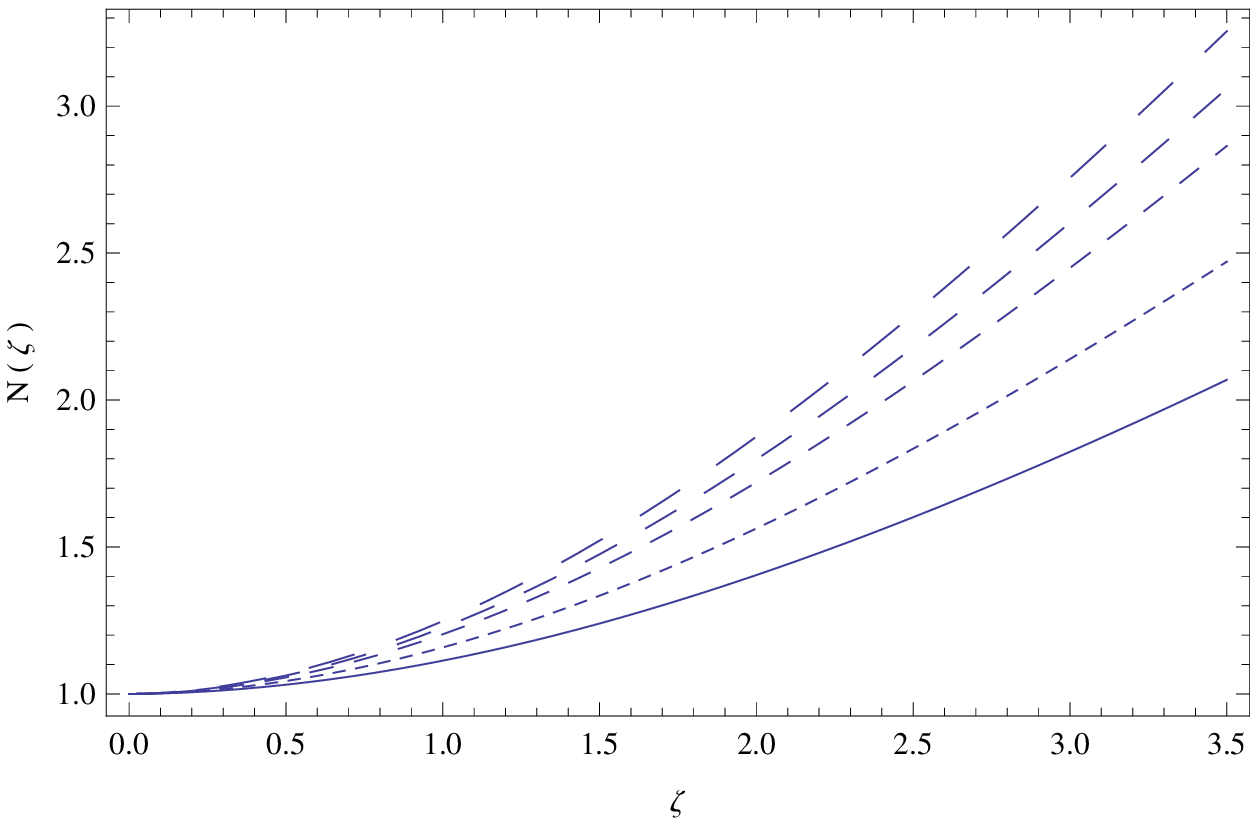}
\end{figure}

\begin{figure}[h]
\caption{Variation of  $L(\zeta) = \sqrt{-g_{\phi
\phi}}$ for the space-time of the quantum pressure dominated
Bose-Einstein condensate string, for different values of $\lambda _Q$%
: $\lambda _Q=0.10$ (solid curve), $\lambda _Q=0.14$ (dotted
curve), $\lambda _Q=0.18$ (short dashed curve), $\lambda %
_Q=0.20 $ (dashed curve), and $\lambda _Q=0.22$ (long dashed curve),
respectively. }
\label{fig12}\centering
\includegraphics[width=8cm]{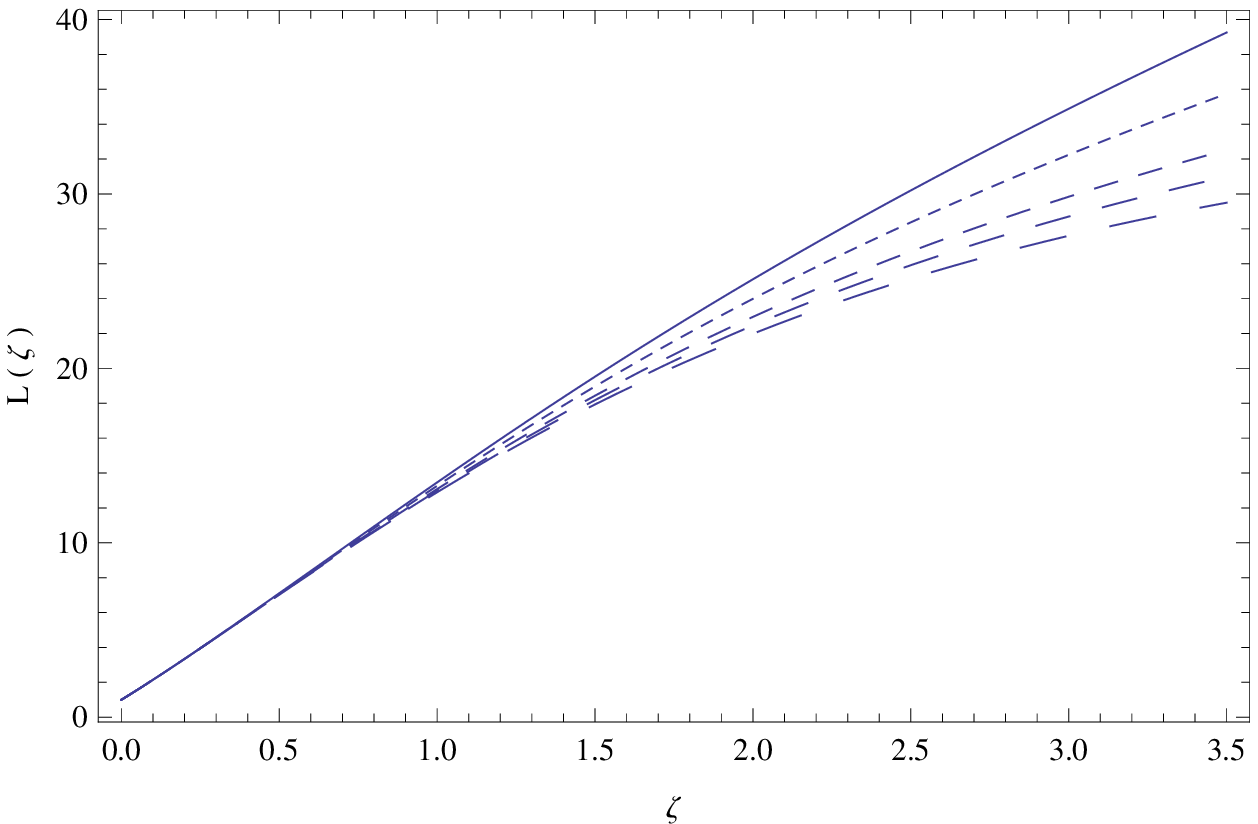}
\end{figure}

\begin{figure}[h]
\caption{Variation of  $K(\zeta) = \sqrt{-g_{zz}}$ for
the space-time of the quantum pressure dominated Bose-Einstein condensate
string, for different values of $\lambda _Q$: $\lambda %
_Q=0.10$ (solid curve), $\lambda _Q=0.14$ (dotted curve), $%
\lambda _Q=0.18$ (short dashed curve), $\lambda _Q=0.20 $ (dashed
curve), and $\lambda _Q=0.22$ (long dashed curve), respectively. }
\label{fig13}\centering
\includegraphics[width=8cm]{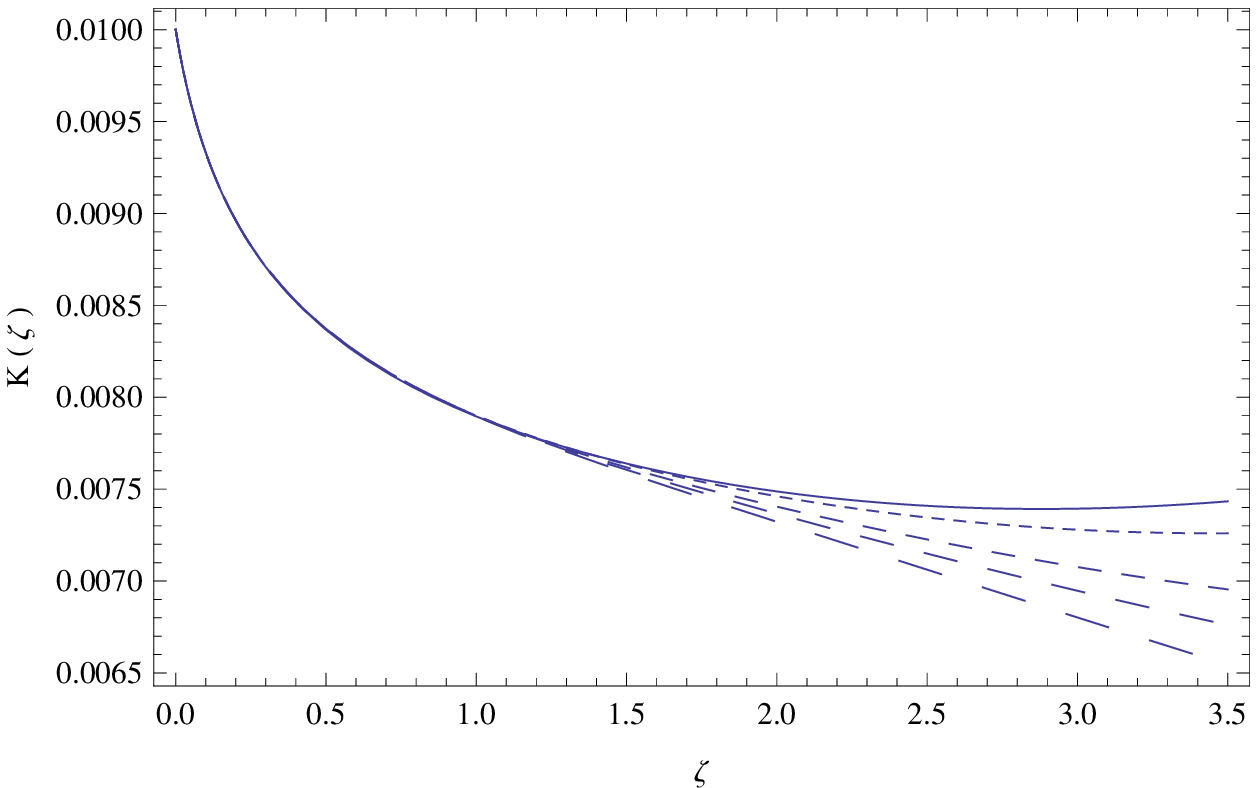}
\end{figure}

\begin{figure}[h]
\caption{Variation of the Tolman mass $m$, as a function of $\zeta$,
of the quantum pressure dominated Bose-Einstein condensate string, for
different values of $\lambda _Q$: $\lambda _Q=0.10$ (solid
curve), $\lambda _Q=0.14$ (dotted curve), $\lambda _Q=0.18$
(short dashed curve), $\lambda _Q=0.20 $ (dashed curve), and $%
\lambda _Q=0.22$ (long dashed curve), respectively. }
\label{fig14}\centering
\includegraphics[width=8cm]{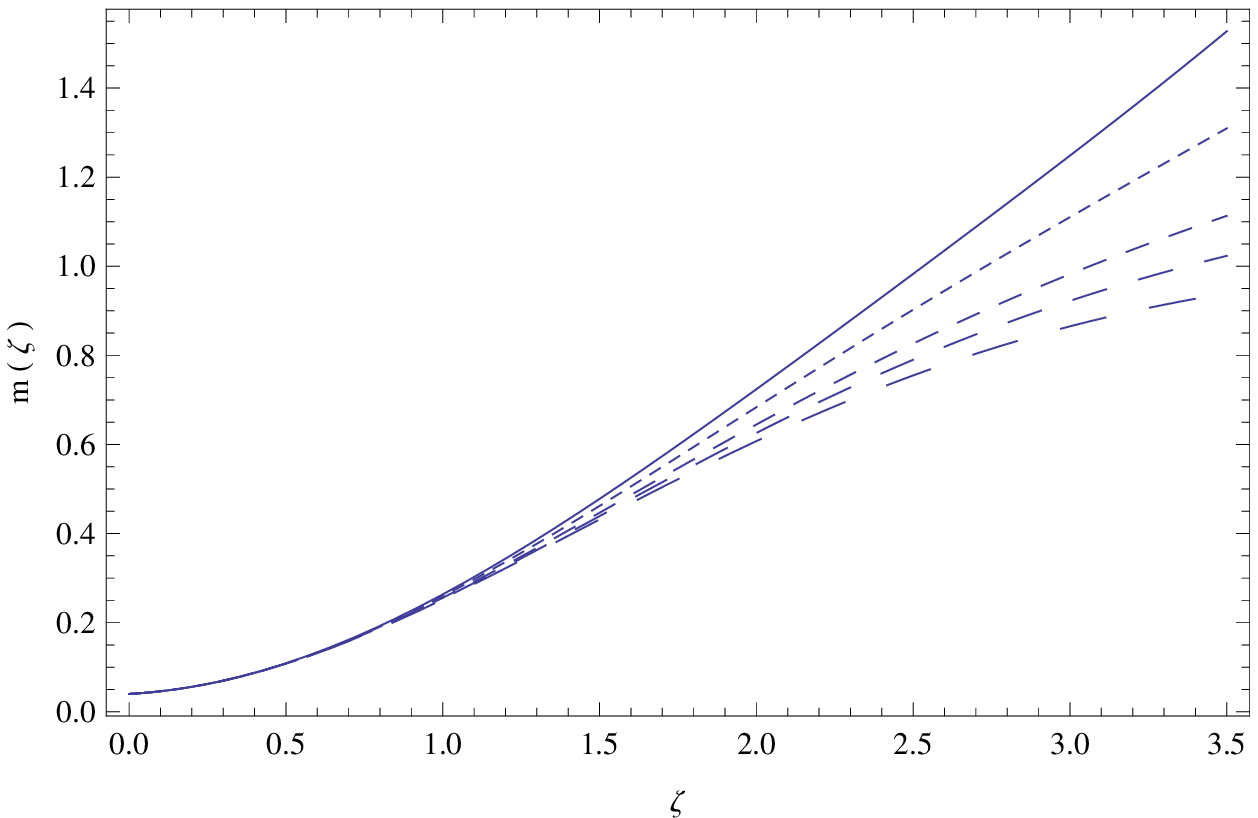}
\end{figure}

\begin{figure}[h]
\caption{Variation of the angular deficit $w$, as a function of $%
\zeta$, of the quantum pressure dominated Bose-Einstein condensate string,
for different values of $\lambda _Q$: $\lambda _Q=0.10$
(solid curve), $\lambda _Q=0.14$ (dotted curve), $\lambda %
_Q=0.18$ (short dashed curve), $\lambda _Q=0.20 $ (dashed curve),
and $\lambda _Q=0.22$ (long dashed curve), respectively. }
\label{fig141}\centering
\includegraphics[width=8cm]{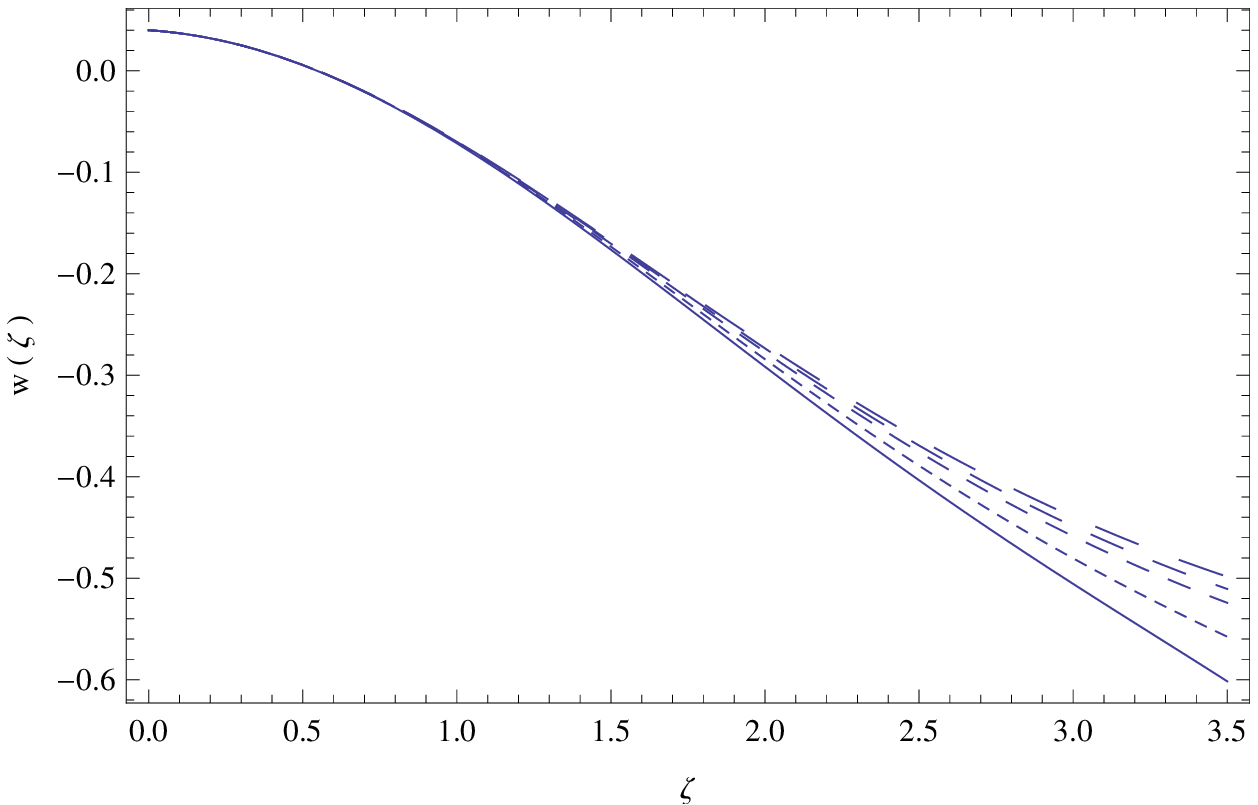}
\end{figure}

The variation, with respect to $\zeta$, of $\Sigma = \sqrt{-g}$, of the
dimensionless energy density of the BEC matter $\tau$, the dimensionless
quantum pressure $p_Q$, the functions $N$, $L$ and $K$ that determine
the components of the metric tensor, the dimensionless Tolman mass $m$,
and the angular deficit parameter $w$ are represented, for different
values of the parameter $\lambda _Q$, in Figs.~\ref{fig8}-\ref{fig141}. In
each case, the initial conditions used to numerically integrate the system
of differential equations, Eqs.~(\ref{iq1})--(\ref{iq4}), were $%
\Sigma(0)=0.01$, $\sigma(0)=0.10$, $\tau(0)=1$, $a(0)=0$, $N(0)=1$, $%
n(0)=-10^{-8}$, $u(0)=-1$, $K(0)=0.01$, $k(0)=-0.01$, $m(0)=0$ and $w(0)=0$.%

The variation of the function $\Sigma(\zeta) = \sqrt{-g}$, presented in Fig~%
\ref{fig8}, appears to be approximately proportional to $\zeta$, for small $%
\zeta$, but varies according to a higher power of $\zeta$ as $\zeta $
increases. This is in contrast to the interaction pressure dominated case,
as well to the $\Sigma = \sqrt{-g} \propto \zeta$ case for a flat
conical geometry. The dimensionless density $\tau(\zeta) =
\rho(\zeta)/\rho_c(\zeta)$ of the BEC cosmic string, plotted in Fig.~\ref%
{fig9}, monotonically decreases from a maximum central value at $\zeta=0$,
reaching the value zero for $\zeta =\zeta _s$, which defines the vacuum
boundary of the string.

For the parameters used in the numerical simulations it follows that $%
R_s\approx 3.5\hbar /2mc=6.148\times 10^{-38}/m$. If the mass of the
particle in the condensate is very small, $m\approx 10^{-44}$ g, then $%
R_s\approx 6.16\times 10^6$ cm. On the other hand, more massive particles
can form condensate strings with very small radii. At the vacuum boundary
$\zeta =\zeta _s$ the quantum pressure, depicted in Fig.~\ref%
{fig10}, tends to zero, together with the energy density of the BEC
particles. However, its relationship with the energy density is now more
complex than in the interaction pressure dominated case. $N(\zeta)/N_0
\propto \sqrt{g_{tt}}$, presented in Fig.~\ref{fig11}, is a monotonically
increasing function of $\zeta$ and it becomes proportional to $\zeta$ at
large radial distances. Its behavior is qualitatively similar to that found
in the space-time of the interaction pressure dominated string. Likewise, $%
L(\zeta) = \sqrt{-g_{\phi \phi}}$, shown in Fig.~\ref{fig12}, behaves in
much the same way as for the interaction pressure dominated string. The
major difference in the behavior of the metric components, between the
interaction and quantum pressure dominated regimes, comes from $K = \sqrt{%
-g_{zz}}$, represented in Fig.~\ref{fig13}. In the latter, $K(\zeta)$
decreases monotonically as a function of the radial distance, with a sharp
immediate drop close to $\zeta = 0$, followed by a more gradual decline.
Again, the Tolman mass function, plotted in Fig.~\ref{fig14}, is
monotonically increasing and tends to a constant value $M\approx
0.8-1.4\times \left(\pi \hbar ^2/2m^2c^2\right)\rho _c\approx 2.71\times
10^{-75}\left(\rho _c/m^2\right)$, giving the total mass of the quantum BEC
string. For an ultralight particle with $m=10^{-44}$ g we obtain $M\approx
2.71\times 10^{13}\rho _c$. Particles with such masses may be the axions of
quantum chromodynamics, pseudo Nambu-Goldstone bosons associated with the
spontaneous breaking of Peccei-Quinn symmetry, which represent an
important dark matter candidate \cite{axion}. Such axions are also believed to form nonstring
BECs \cite{Sikivie:2009qn} and, while astrophysical constraints on axionlike dark matter candidates can be obtained
from big bang nucleosynthesis data, a number of direct detection schemes have recently been proposed, based 
on their predicted interactions with solid state and atomic systems  \cite{Budker:2013hfa,Roberts:2014dda}.
By using the most recent cosmological data, including the Planck temperature data, the WMAP
E-polarization measurements, the recent BICEP2 observations of B-modes, as
well as baryon acoustic oscillation data, including those from the Baryon
Oscillation Spectroscopic Survey \cite{Valent}, it was found that the
mass of dark matter axions is constrained to lie in the range of $70-80$ $\mu\mathrm{eV}$.
Lighter dark matter particles are also possible.

Thus, if the central density is sufficiently high, quantum strings may have masses
of the order of $M\approx 10^{-7}M_{\odot}$. The angular deficit $w$, shown
in Fig.~\ref{fig141}, is again negative inside the string, and it is a
monotonically decreasing function of the radial distance. This suggests that
the possibility of BEC strings giving rise to space-time angle excess
cannot be excluded in either the $p \gg P_Q$ or $p \ll P_Q$ regimes and,
hence, that it may, in principle, be a generic feature of such strings,
given appropriate initial conditions. However, further investigation is
required in order to determine just how generic these conditions may be, and
whether or not they are comparable in each regime.

\subsection{Quantum pressure dominated Bose-Einstein condensate strings: the
effects of the gravitational field on the equation of state}\label{new}

In the previous section, we considered the properties of quantum
pressure dominated Bose-Einstein condensate strings by assuming that the
equation of state of the string is independent of the background
gravitational field. From a mathematical point of view, this means we have
assumed that the Laplacian operator $\vec{\nabla}^2$, acting on the density, can
be defined in the standard Euclidian space of quantum mechanics. Hence, in
this approach, the local quantum equation of state is not affected by
the curvature of the space-time (the gravitational field), an
approximation which is consistent with the semiclassical limit of general
relativity. However, since the quantum equation of state also contains a
geometric component (via the presence of the geometry-dependent Laplacian),
in the presence of strong gravitational fields, the geometry of
the space-time may play a significant role in the physical description of
the quantum pressure. Therefore, in this section, we consider the
effects of the gravitational field on the quantum pressure equation of
state. In order to obtain the equation of state in a curved
space-time, we introduce the corresponding three-dimensional Laplacian as
\begin{equation}
\vec{\nabla}^2 =\frac{1}{L(r)K(r)}\frac{d}{dr}\left[ L(r)K(r)\frac{d}{dr}\right] ,
\end{equation}%
which allows us to define the quantum pressure as
\begin{equation}
p_{Q}(r)=-\frac{\hbar ^{2}}{4m^{2}}\rho (r)\frac{1}{L(r)K(r)}\frac{d}{dr}%
\left[ L(r)K(r)\frac{d}{dr}\ln \frac{\rho (r)}{\rho _{c}}\right] .
\end{equation}

Hence, by taking into account the gravitational effects on the quantum
equation of state, the field equations describing the quantum pressure
supported BEC string take the form
\begin{equation}
\frac{1}{\Sigma }\frac{d}{dr}\left( \Sigma H_{t}\right) =\frac{4\pi G}{c^{2}}%
\rho \left[ 1-3\frac{\hbar ^{2}}{4m^{2}c^{2}}\frac{1}{LK}\frac{d}{dr}\left(
LK\frac{d}{dr}\ln \frac{\rho }{\rho _{c}}\right) \right] ,
\end{equation}%
\begin{eqnarray}
&&3\frac{dH}{dr}+H_{t}^{2}+H_{\phi }^{2}+H_{z}^{2}=  \nonumber \\
&&-\frac{4\pi G}{c^{2}}\rho \left[ 1+\frac{\hbar ^{2}}{4m^{2}c^{2}}\frac{1}{%
LK}\frac{d}{dr}\left( LK\frac{d}{dr}\ln \frac{\rho }{\rho _{c}}\right) %
\right] ,
\end{eqnarray}%
\begin{eqnarray}
\frac{1}{\Sigma }\frac{d}{dr}\left( \Sigma H_{i}\right) &=&-\frac{4\pi G}{%
c^{2}}\rho \left[ 1+\frac{\hbar ^{2}}{4m^{2}c^{2}}\frac{1}{LK}\frac{d}{dr}%
\left( LK\frac{d}{dr}\ln \frac{\rho }{\rho _{c}}\right) \right] ,  \nonumber
\\
&&i=\phi ,z,
\end{eqnarray}%
where $\rho _{c}$ is again the central density of the string. The corresponding
Tolman mass is given by
\bea
M(r)&=&2\pi \int_{r'=0}^{r}\rho \left(r'\right)\Bigg\{ 1+3\frac{\hbar }{4m^{2}c^{2}}\frac{1}{L\left(r'\right)K\left(r'\right)}\times \nonumber\\
&&\frac{d}{dr'}\left[ L\left(r'\right)K\left(r'\right)\frac{d}{dr'}\ln \frac{\rho \left(r'\right)}{\rho _{c}}\right] \Bigg\}
\Sigma\left(r'\right)dr',\nonumber\\
\eea%
while
\bea
&&W(r)=-2\pi \int_{r'=0}^{r}\rho \left(r'\right)\Bigg\{ 1+\frac{\hbar }{4m^{2}c^{2}}\frac{1}{L\left(r'\right)K\left(r'\right)}\times \nonumber\\
&&\frac{d}{dr'}\left[ L\left(r'\right)K\left(r'\right)\frac{d}{dr'}\ln \frac{\rho \left(r'\right)}{\rho _{c}}\right] \Bigg\}
\Sigma\left(r'\right) dr'.
\eea

The relation between $H_{\phi }$ and $H_{z}$ is again obtained as $H_{\phi
}=H_{z}+C/\Sigma $. By introducing the same set of dimensionless quantities
as in the previous section, we obtain the basic equations of the new model in a dimensionless form:
\begin{equation}\label{n1}
\frac{1}{\Sigma }\frac{d^{2}\Sigma }{d\zeta ^{2}}=-\lambda _{Q}\tau \left[
1+5\frac{1}{LK}\frac{d}{d\zeta }\left( LK\frac{d}{d\zeta }\ln \tau \right) %
\right] ,
\end{equation}
\bea\label{n2}
&&\frac{d}{d\zeta }\left[ \tau \frac{1}{LK}\frac{d}{d\zeta }\left( LK\frac{d}{%
d\zeta }\ln \tau \right) \right] -\tau \Bigg[ 1-\nonumber\\
&&\frac{1}{LK}\frac{d}{d\zeta }%
\left( LK\frac{d}{d\zeta }\ln \tau \right) \Bigg] {\mathcal{H}}_{t}=0,
\eea
\begin{equation}\label{n3}
\frac{d{\mathcal{H}}_{t}}{d\zeta }+\frac{1}{\Sigma }\frac{d\Sigma }{d\zeta }{%
\mathcal{H}}_{t}=\lambda _{Q}\tau \left[ 1-3\frac{1}{LK}\frac{d}{d\zeta }%
\left( LK\frac{d}{d\zeta }\ln \tau \right) \right] ,
\end{equation}%
\begin{equation}\label{n4}
\frac{d{\mathcal{H}}_{\phi }}{d\zeta }+\frac{1}{\Sigma }\frac{d\Sigma }{%
d\zeta }{\mathcal{H}}_{\phi }=-\lambda _{Q}\tau \left[ 1+\frac{1}{LK}\frac{d%
}{d\zeta }\left( LK\frac{d}{d\zeta }\ln \tau \right) \right] ,
\end{equation}
\be\label{n5}
\frac{d{\mathcal{H}}_{z}}{d\zeta }+\frac{1}{\Sigma }\frac{d\Sigma }{d\zeta }{%
\mathcal{H}}_{z}=-\lambda _{Q}\tau \left[ 1+\frac{1}{LK}\frac{d}{d\zeta }%
\left( LK\frac{d}{d\zeta }\ln \tau \right) \right] ,
\ee
\begin{equation}\label{n6}
m(\zeta )=2\pi \int_{\zeta' =0}^{\zeta }{\tau \left[ 1+3\frac{1}{LK}\frac{d}{%
d\zeta' }\left( LK\frac{d}{d\zeta' }\ln \tau \right) \right] \Sigma d\zeta' },
\end{equation}
\begin{equation}\label{n7}
w(\zeta )=-2\pi \int_{\zeta' =0}^{\zeta }{\tau \left[ 1+\frac{1}{LK}\frac{d}{%
d\zeta' }\left( LK\frac{d}{d\zeta' }\ln \tau \right) \right] \Sigma d\zeta' }.
\end{equation}

The quantum pressure can be obtained in a dimensionless form as
\be
P_Q=-\tau \frac{1}{LK}\frac{d}{d\zeta }\left(LK\frac{d}{d\zeta }\ln \tau \right).
\ee

\begin{figure}[h]
\caption{Variation of $\Sigma = \sqrt{-g}$, as a function of $%
\zeta$, in the space-time of the quantum pressure supported Bose-Einstein condensate string with geometry-dependent equation of state, for different values of $\lambda _Q$: $%
\lambda _Q=0.10$ (solid curve), $\lambda _Q=0.14$ (dotted
curve), $\lambda _Q=0.18$ (short dashed curve), $\lambda %
_Q=0.20 $ (dashed curve), and $\lambda _Q=0.22$ (long dashed curve),
respectively. }
\label{fig15}\centering
\includegraphics[width=8cm]{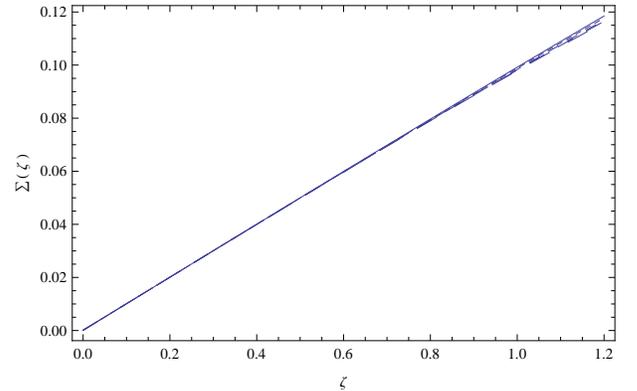}
\end{figure}

\begin{figure}[h]
\caption{Variation of the dimensionless density $\tau $, as a
function of $\zeta$, of the quantum pressure dominated Bose-Einstein
condensate string with geometry-dependent equation of state, for different values of $\lambda _Q$: $%
\lambda _Q=0.10$ (solid curve), $\lambda _Q=0.14$ (dotted curve), $%
\lambda _Q=0.18$ (short dashed curve), $\lambda _Q=0.20 $
(dashed curve), and $\lambda _Q=0.22$ (long dashed curve),
respectively. }
\label{fig16}\centering
\includegraphics[width=8cm]{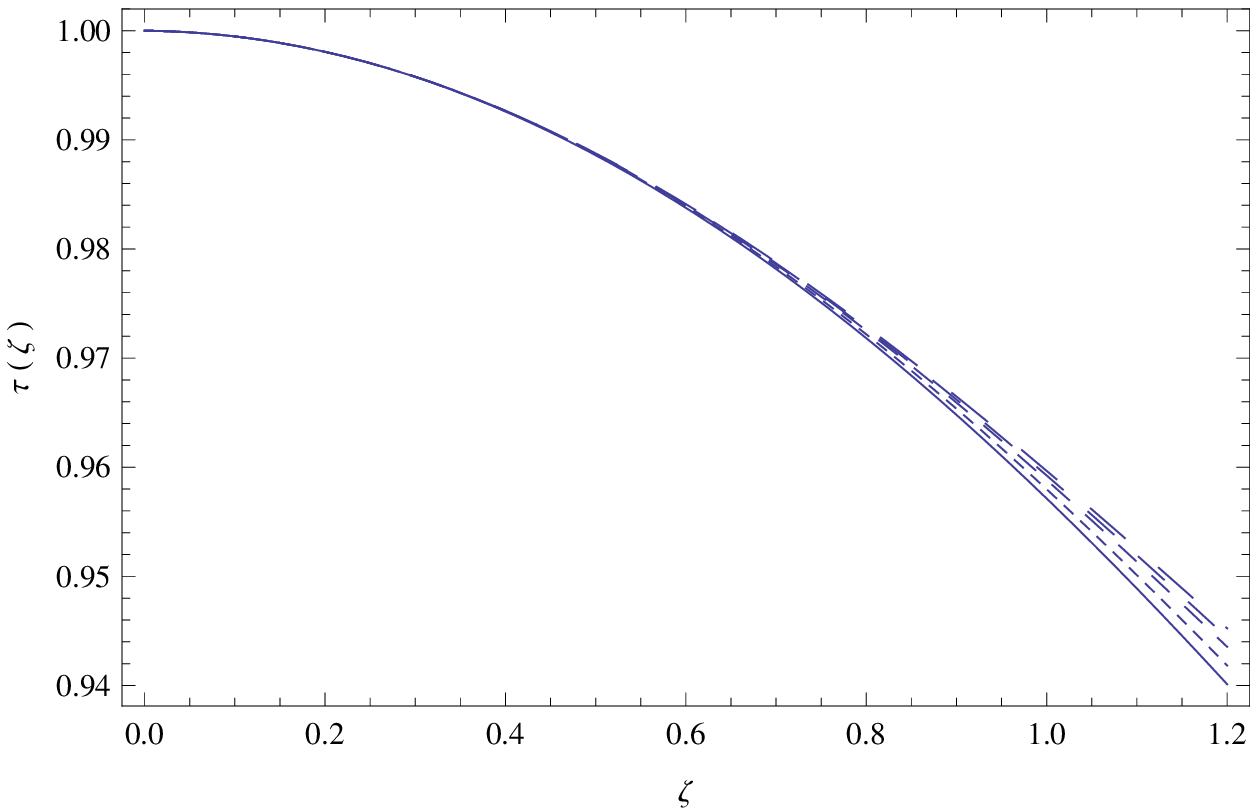}
\end{figure}

\begin{figure}[h]
\caption{Variation of the dimensionless quantum pressure $P_Q$, as a
function of $\zeta$, of the quantum pressure dominated Bose-Einstein
condensate string with geometry-dependent equation of state, for different values of $\lambda _Q$: $%
\lambda _Q=0.10$ (solid curve), $\lambda _Q=0.14$ (dotted curve), $%
\lambda _Q=0.18$ (short dashed curve), $\lambda _Q=0.20 $
(dashed curve), and $\lambda _Q=0.22$ (long dashed curve),
respectively. }
\label{fig17}\centering
\includegraphics[width=8cm]{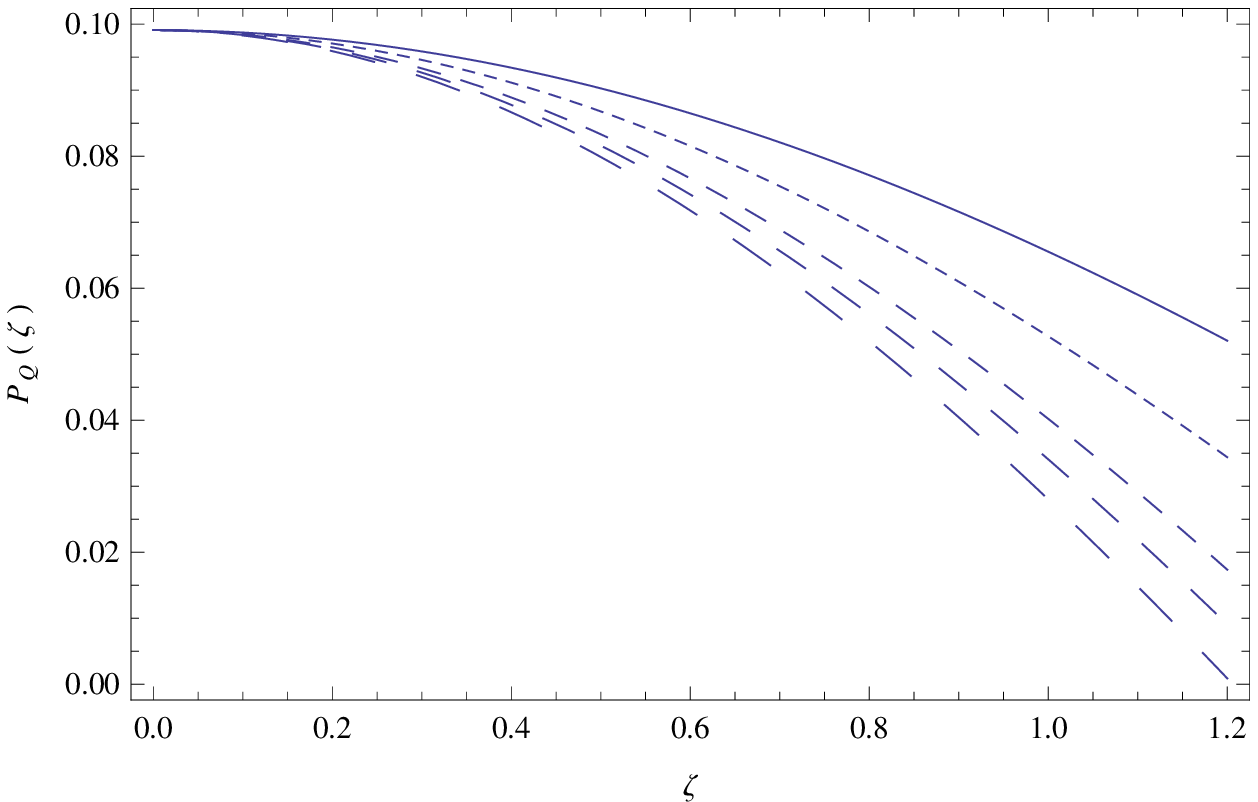}
\end{figure}

\begin{figure}[h]
\caption{Variation of the $N(\zeta) = \sqrt{g_{tt}}$ for the
space-time of the quantum pressure dominated Bose-Einstein condensate string with geometry-dependent equation of state,
for different values of $\lambda _Q$: $\lambda _Q=0.10$
(solid curve), $\lambda _Q=0.14$ (dotted curve), $\lambda %
_Q=0.18$ (short dashed curve), $\lambda _Q=0.20 $ (dashed curve),
and $\lambda _Q=0.22$ (long dashed curve), respectively. }
\label{fig18}\centering
\includegraphics[width=8cm]{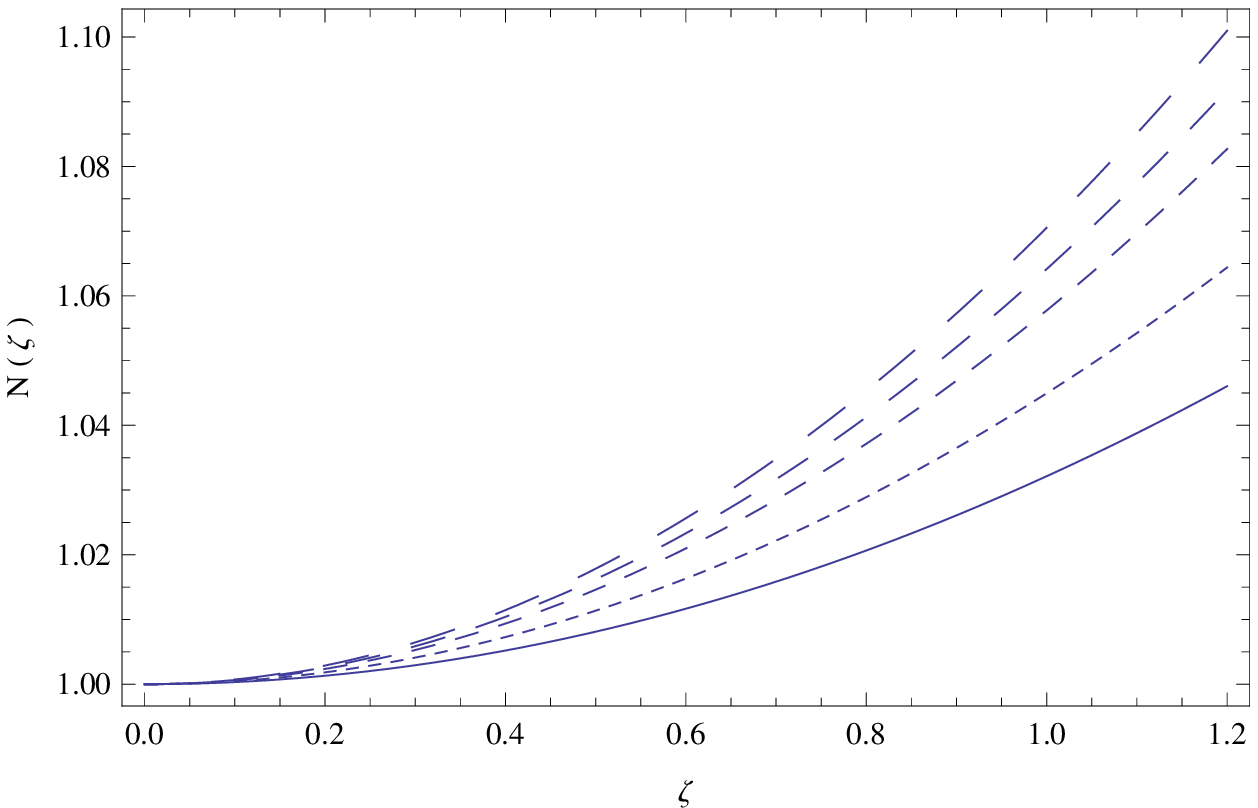}
\end{figure}

\begin{figure}[h]
\caption{Variation of  $L(\zeta) = \sqrt{-g_{\phi
\phi}}$ for the space-time of the quantum pressure dominated
Bose-Einstein condensate string with geometry-dependent equation of state, for different values of $\lambda _Q$%
: $\lambda _Q=0.10$ (solid curve), $\lambda _Q=0.14$ (dotted
curve), $\lambda _Q=0.18$ (short dashed curve), $\lambda %
_Q=0.20 $ (dashed curve), and $\lambda _Q=0.22$ (long dashed curve),
respectively. }
\label{fig19}\centering
\includegraphics[width=8cm]{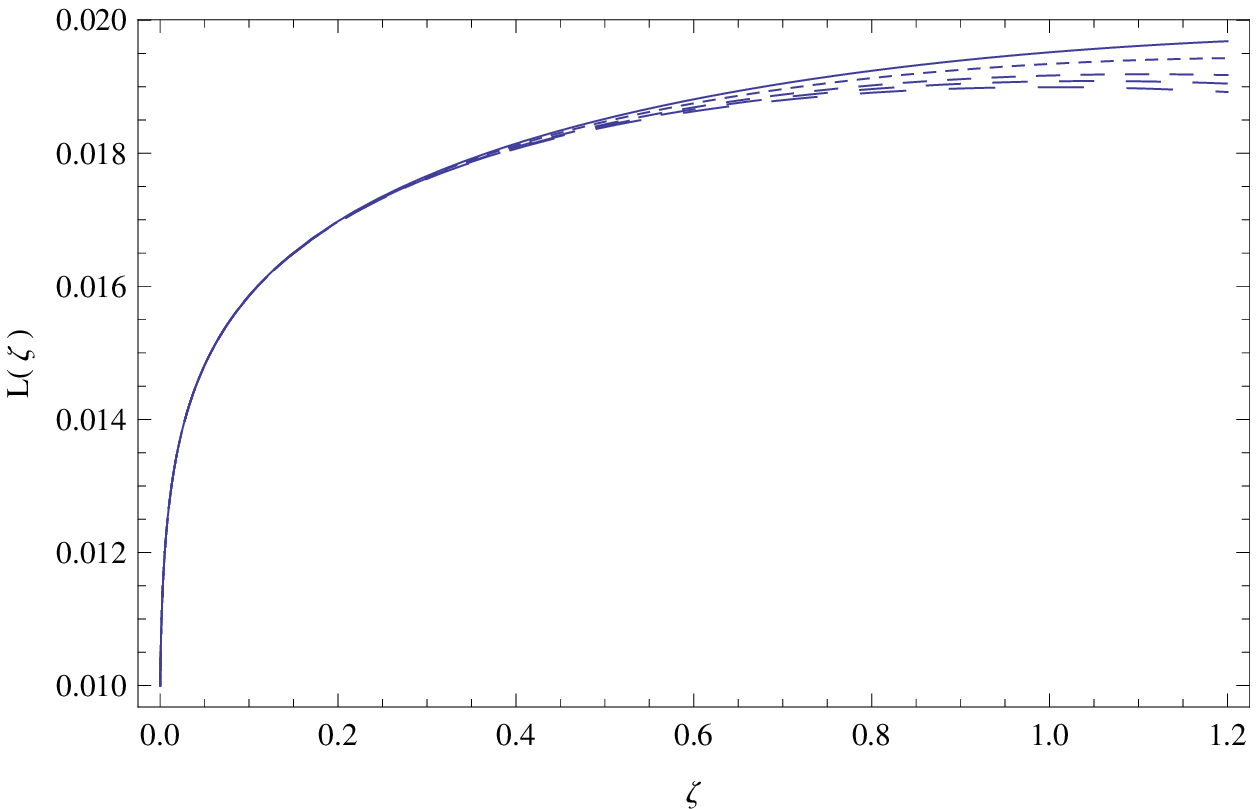}
\end{figure}

\begin{figure}[h]
\caption{Variation of  $K(\zeta) = \sqrt{-g_{zz}}$ for
the space-time of the quantum pressure dominated Bose-Einstein condensate
string with geometry-dependent equation of state, for different values of $\lambda _Q$: $\lambda %
_Q=0.10$ (solid curve), $\lambda _Q=0.14$ (dotted curve), $%
\lambda _Q=0.18$ (short dashed curve), $\lambda _Q=0.20 $ (dashed
curve), and $\lambda _Q=0.22$ (long dashed curve), respectively. }
\label{fig20}\centering
\includegraphics[width=8cm]{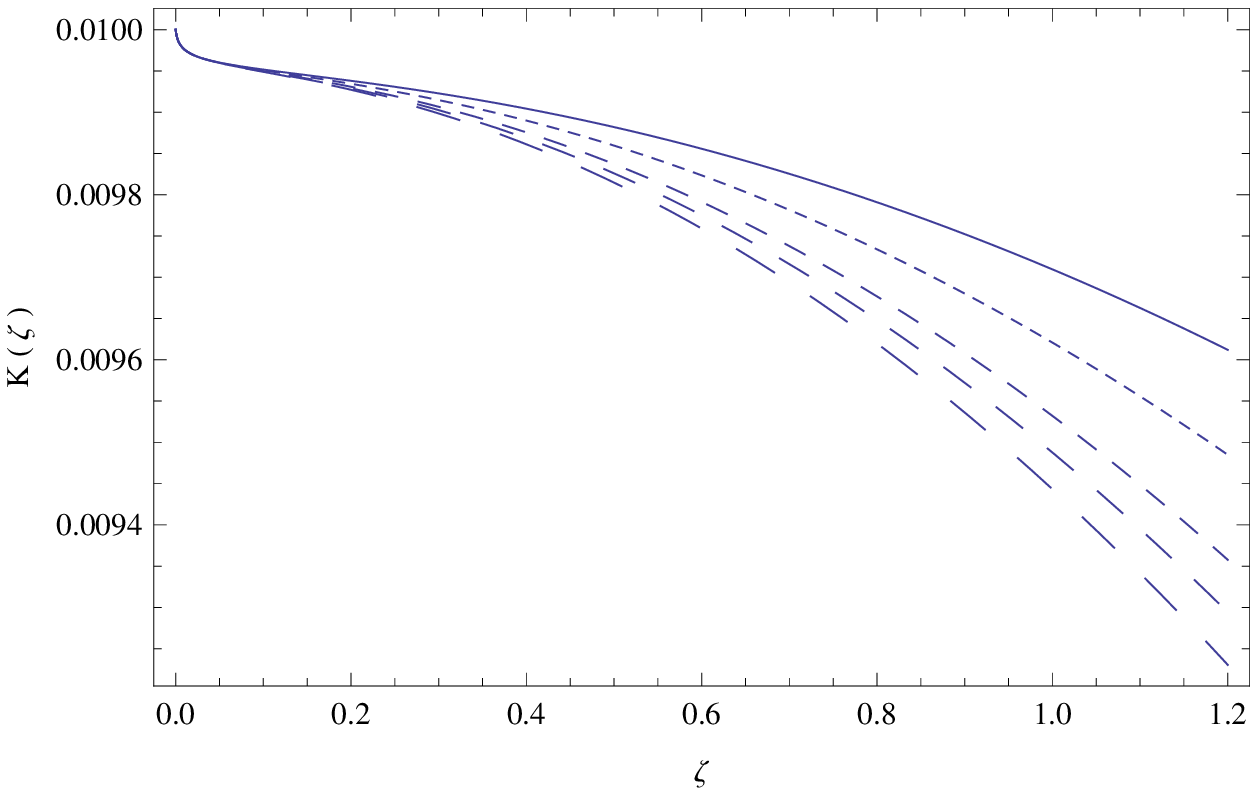}
\end{figure}

\begin{figure}[h]
\caption{Variation of the Tolman mass $m$, as a function of $\zeta$,
of the quantum pressure dominated Bose-Einstein condensate string with geometry-dependent equation of state, for
different values of $\lambda _Q$: $\lambda _Q=0.10$ (solid
curve), $\lambda _Q=0.14$ (dotted curve), $\lambda _Q=0.18$
(short dashed curve), $\lambda _Q=0.20 $ (dashed curve), and $%
\lambda _Q=0.22$ (long dashed curve), respectively. }
\label{fig21}\centering
\includegraphics[width=8cm]{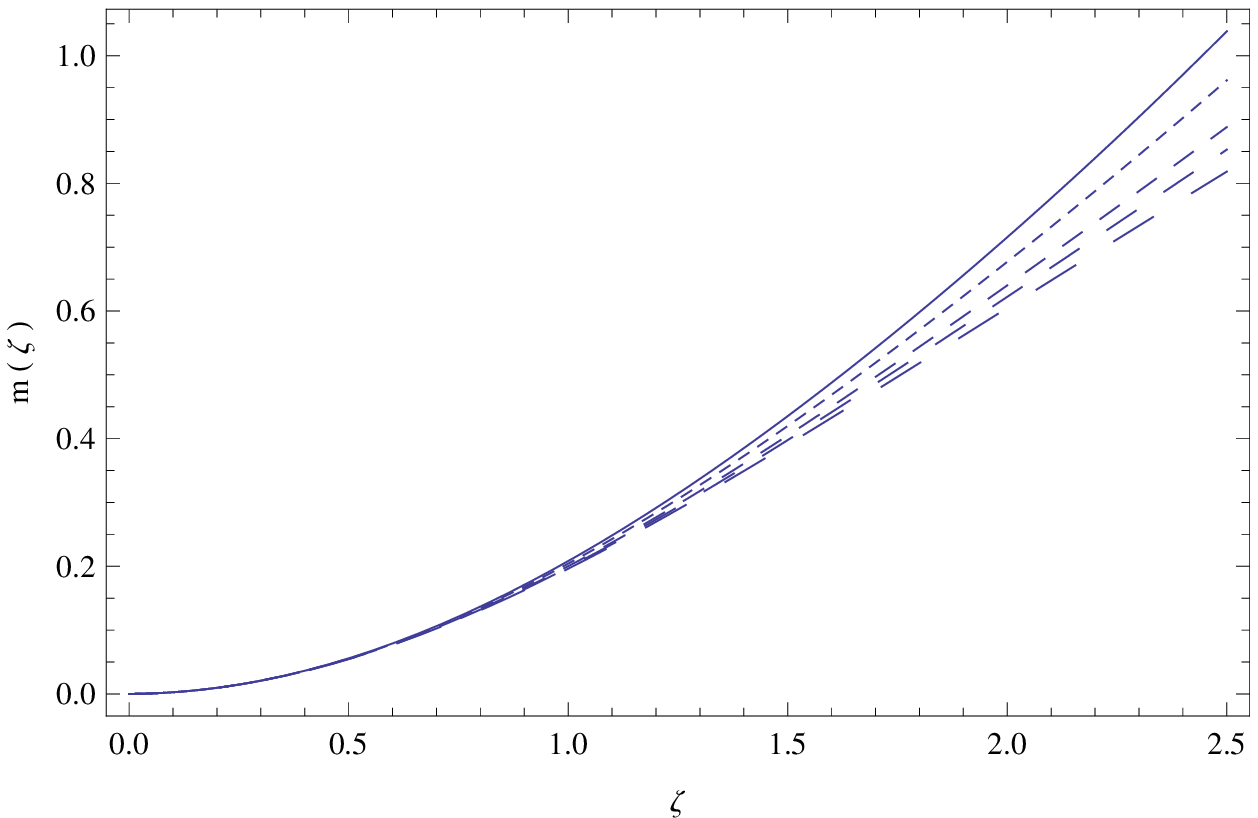}
\end{figure}

\begin{figure}[h]
\caption{Variation of the angular deficit $w$, as a function of $%
\zeta$, of the quantum pressure dominated Bose-Einstein condensate string with geometry-dependent equation of state,
for different values of $\lambda _Q$: $\lambda _Q=0.10$
(solid curve), $\lambda _Q=0.14$ (dotted curve), $\lambda %
_Q=0.18$ (short dashed curve), $\lambda _Q=0.20 $ (dashed curve),
and $\lambda _Q=0.22$ (long dashed curve), respectively. }
\label{fig22}\centering
\includegraphics[width=8cm]{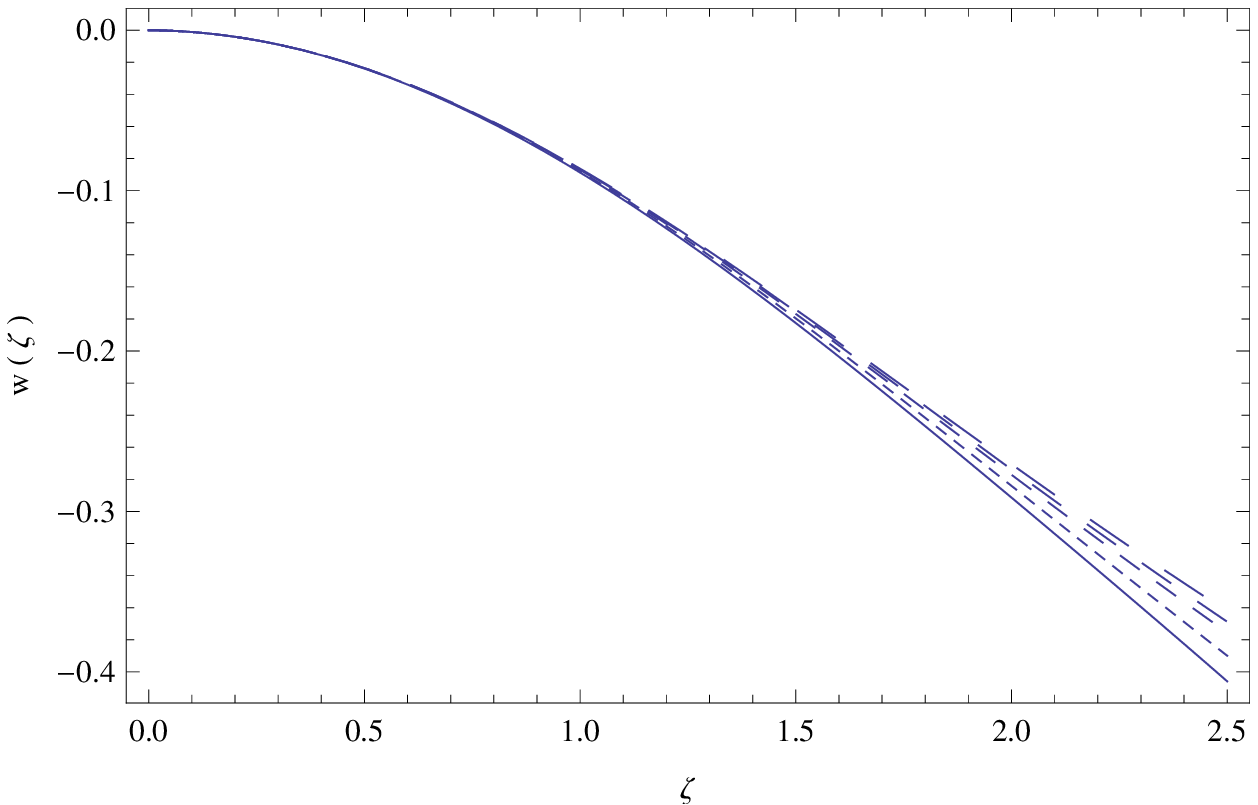}
\end{figure}

In order to numerically integrate the system of Eqs.~(\ref{n1})-(\ref{n7}), we choose the initial conditions identical to those in the previous section. Hence we adopt the initial values for the geometrical and physical parameters on the string axis as  $N(0)=1$, $L(0)=0.01$, $K(0)=0.01$, $H_t(0)=N'(0)/N(0)=-10^(-8)$, $H_{\phi}(0)=L'(0)/L(0)=100$, $H_z(0)=K'(0)/K(0)=-1$, $\tau (0)=1$, $\tau '(0)=0$, $\tau ''(0)=-0.0001$, $\Sigma (0)=10^{-8}$, $\Sigma '(0)=0.1$, $m(0)=0$ and $w(0)=0$, respectively. Hence we have also adopted identical initial conditions for both $L$ and $K$. The variations, with respect to the dimensionless radial distance $\zeta $, of the square root of the determinant of the metric tensor $\Sigma = NKL$, the individual functions $N$, $L$ and $K$, the dimensionless mass density $\tau $, and the quantum pressure inside the string $P_Q$, as well as the Tolman mass and the angular deficit parameter $w$, are presented, for different values of the free parameter $\lambda _Q$, in Figs.~\ref{fig15}-\ref{fig22}.

The behavior of the physical parameters of the Bose-Einstein condensate string, in the presence of a space-time geometry-dependent quantum pressure, is qualitatively similar to the behavior of the same parameters in the case of the ``simplified" quantum equation of state, in which the influence of the background geometry is neglected. However, some quantitative differences do arise due to the gravitational influence on the equation of state. As a general result,
and for the specific initial values of the geometrical and physical quantities adopted in the examples given above, the string becomes more compact, and its radius as well as its mass decrease slightly, as compared to semiclassical approximation. Again, the square root of the determinant of the metric tensor, shown in Fig.~\ref{fig15}, is a monotonically increasing function of the radial distance $\zeta$. For small values of $\zeta $ the increase is linear, and practically independent of the values of the parameter $\lambda _Q$. For large values of $\zeta $ the increase can be approximately described by a second degree algebraic function, but the deviations from the linear regime are extremely small. The energy density $\tau $ of the string, represented in Fig.~\ref{fig16}, is a monotonically decreasing function of $\zeta$. As usual, the quantum pressure, presented in Fig.~\ref{fig17}, monotonically decreases with $\zeta $ and reaches zero at the vacuum boundary of the string, corresponding to $P_Q \left(R_s\right)=0$. The geometric effects on the equation of state significantly reduce the central values of the quantum pressure, due to the explicit dependence on the metric functions,  while increasing the rate at which $P_Q$ approaches zero. The value of the radial coordinate $\zeta $ for which the quantum pressure becomes zero can be taken as describing the quantum string radius, which for the adopted values of the parameter $\lambda _Q$ is in the range  $R_s\approx (1.2-2)\times \hbar /2mc=(2.108-3.513)\times 10^6\times \left(m/10^{-44}\;{\rm g}\right)^{-1}\;{\rm cm}$. For a particle with mass $m=10^{-44}$ g the Bose-Einstein condensate string radii are  of the same order of magnitude as the neutron star radii. This radius is slightly lower than the corresponding radius of the simplified quantum string model, showing that gravitational effects in the quantum pressure indeed make the string more compact.

The metric tensor components $N(\zeta)$ and  $L(\zeta)$, represented in Figs.~\ref{fig18} - \ref{fig19}, are increasing functions of $\zeta $, and their behavior is similar to in the simplified quantum pressure case. For small values of $\zeta$, $L(\zeta )$ increases linearly, but sharply, while $N(\zeta )$ is roughly constant. For large values of $\zeta $, near the string vacuum boundary, $N(\zeta)$ has an approximate second-order polynomial behavior, while $L(\zeta)$ has approximately constant values. Similar to the case of the quantum pressure supported string in the semiclassical approximation, the $g_{zz}$ metric tensor component 
$K(\zeta)$, shown in Fig.~\ref{fig20}, is a decreasing function of $\zeta$. For small values of $\zeta $, the behavior of all three metric functions is independent of $\lambda _Q$, and can be approximated by a linear function in the radial distance $\zeta $. The Tolman mass $m$, represented in Fig.~\ref{fig21}, is a monotonically increasing function, and reaches the value $m\left(\zeta _s\right)\approx 0.35-0.40$ at the string surface. The total mass of the string can therefore be estimated as $M\approx (6.786-7.755)\times 10^{12}\times \left(m/10^{-44}\;{\rm g}\right)^{-2}\times \rho _c\;{\rm g}$, which is slightly lower, as compared to the mass of the string in the semiclassical model, in which the effect of the geometry on the quantum equation of state is neglected. The angular deficit parameter inside the string $w$ , plotted in Fig.~\ref{fig22}, is again negative, with values in the range $w\in (0, -0.4)$, and reaches higher values on the string boundary, as compared to the simplified, geometry-independent case.

\section{The Newtonian Approximation}

\label{Sect.IV}

Due to the invariance of the action of the BEC system, Eq.~(\ref{ham}),
under infinitesimal time translations $t\rightarrow t+\delta $ (with $\delta
\vec{r}=\delta \psi =\delta \psi ^{*}=0$), the nonrelativistic Hamiltonian,
and hence the total energy of the condensed particles in the semiclassical
approach to quantum BEC strings in general relativity, is conserved.
However, in the Newtonian approximation, the total energy $E$ of a
gravitationally bound BEC can, in general, be written in the especially
simple way, $E = E_{kin} + E_{int} + E_{grav}$, where $E_{kin}$, $E_{int}$
and $E_{grav}$ denote the kinetic, interaction and gravitational energy,
respectively. In this final section, we consider an approximate estimate of
the energy only, based on the method used in \cite{Pet}, which assumes the
Newtonian limit for the gravitational field. The kinetic energy per particle
is of order $\sim (\hbar ^2/2mR^2 + \hbar^2/2m\Delta^2)$, where $R$ gives
the radial extension and $\Delta$ is the length of the cylinder. Therefore,
the approximate expression for the total kinetic energy of the system is
given by
\begin{equation}
E_{kin} \approx \frac{\mathcal{N}\hbar ^2(R^2+\Delta^2)}{2m R^2\Delta^2}.
\end{equation}
The interaction energy is given by $E_{int} \approx (1/2) (\mathcal{N}^2/\mathcal{V}%
)u_0$, where $\mathcal{V}$ again denotes the volume of the condensate and $\mathcal{N}$
is the total number of particles, while the gravitational potential energy
is $E_{grav} \approx - G\mathcal{M}^2/R$, where $\mathcal{M}$ is the total
mass and, for simplicity, we have neglected the numerical factor 
in the expression for the gravitational potential energy corresponding to a
cylindrically symmetric mass distribution. Therefore, in the Newtonian
approximation, and neglecting numerical factors of order unity, the total
energy of the condensate is given by
\begin{equation}  \label{en}
E \approx \mathcal{N}\frac{\hbar ^2(R^2+\Delta^2)}{mR^2\Delta^2} + \mathcal{N}^2\frac{\hbar^2 l_s%
}{mR^2\Delta}-\frac{G\mathcal{M}^2}{R},
\end{equation}
where we have used the expressions $u_0 \sim \lambda \sim \hbar^2 l_s/m$ and
$\mathcal{V} \sim R^2\Delta$. From Eq.~(\ref{en}) we can obtain a rough
estimate of the physical parameters of the BEC string in the interaction
energy and quantum potential dominated regimes. The quantum kinetic energy
is much bigger than the interaction energy when the parameters of the system
satisfy the condition
\begin{equation}  \label{delta1}
\Delta \gg \mathcal{N} l_s \left[1 + \mathcal{O}\left(\frac{R^2}{\Delta}\right)^2%
\right],
\end{equation}
for $\Delta \gg R$, or
\begin{equation}  \label{R1}
R \gg \mathcal{N} l_s \left[1 + \mathcal{O}\left(\frac{\Delta^2}{R^2}\right)^2\right],
\end{equation}
for $\Delta \ll R$. Both expressions, Eqs. (\ref{delta1})-(\ref{R1}), hold
(approximately), in the limiting case, $\Delta \sim R$, for which the
result for a spherically symmetric BEC matter distribution is recovered \cite%
{Pet}. Assuming the former case, $\Delta \gg R$, which is far more probable
for astrophysical BEC strings, the approximate total energy is given by
\begin{equation}
E \approx \mathcal{N}\frac{\hbar^2}{m R^2}-\frac{G\mathcal{N}^2m^2}{R}\leq 0,
\end{equation}
where we have used $\mathcal{M} \sim m \mathcal{N}$. This, in turn, gives a lower
bound for the radius of a quantum pressure dominated BEC string, i.e.
\begin{equation}  \label{R2}
R_{quant} \gtrsim \frac{\hbar ^2}{Gm^3\mathcal{N}}.
\end{equation}
Equivalently, this may be rewritten as a bound on the (dimensionless)
measure of the mass per unit length, which we now label $G\mu_{quant}$,
using $\mathcal{M} = \mu_{quant}\Delta$:
\begin{equation}  \label{mu1}
G\mu_{quant} \gtrsim \frac{\hbar ^2}{m^2 R\Delta},
\end{equation}
which illustrates the problem, already mentioned in Sec. \ref{Sect.IIIC},
with taking the wire approximation for BEC strings. Due to the presence of
the quantum pressure term, which becomes significant for narrow strings
since only a small number of particles inhabit a thin cross section, the
mass per unit length depends sensitively on the product $R\Delta$. Formally,
we may consider an infinitely long string of zero width by taking the limits
$\Delta \rightarrow \infty$, $R \rightarrow 0$, such that $R\Delta$ remains
finite. Realistically, however, we would like to be able to treat open
strings and loops as well as, on cosmological scales, strings whose width is
limited simply by causality and the finiteness of the horizon. In such
cases, the energy density of the string diverges in the wire approximation
because the surface tension becomes focused on an area that shrinks to
zero. As stated previously, we would therefore expect both finite width
corrections \cite{MaedaTurok(1988)} and effective rigidity terms \cite%
{Anderson et al.(1997),Gregory(1988),Gregory(1988)*} to be important in
constructing an effective action for BEC strings. It is interesting to note
that, for the quantum pressure dominated BEC, neither the minimum string
radius, nor the minimum mass per unit length depends on the scattering length.%
\newline
\indent
By contrast, the interaction energy dominates the internal dynamics of BEC
strings when
\begin{equation}  \label{Delta1}
\Delta \ll \mathcal{N} l_s \left[1 + \mathcal{O}\left(\frac{R^2}{\Delta}\right)^2%
\right],
\end{equation}
assuming $\Delta \gg R$. This condition is obviously satisfied by
cylindrical BEC systems with very large numbers of particle number in
relation to their length or, in other words, by thicker strings (for fixed $%
l_s$), though we cannot say \emph{a priori} whether a string of radius $R$
and length $\Delta$ will be dominated by quantum or interaction pressure.
The condition for stability in the latter case is
\begin{equation}
E \approx \mathcal{N}^2\frac{\hbar ^2l_s}{mR^3}-\frac{Gm^2\mathcal{N}^2}{R}\leq 0,
\end{equation}
which gives the following constraint for the radius, $R_{int}$, of an
interaction pressure dominated string,
\begin{equation}  \label{R3}
R_{int} \gtrsim \sqrt{\frac{\hbar ^2l_s}{Gm^3}}.
\end{equation}
Thus, in this regime, we have that
\begin{equation}  \label{*}
\sqrt{\frac{\hbar ^2l_s}{Gm^3}} \lesssim R \ll \Delta \ll \mathcal{N} l_s,
\end{equation}
from which it follows that interaction pressure dominated strings, with
characteristic minimum radius, form when BEC systems with a sufficient
number of particles, i.e.
\begin{equation}  \label{**}
\mathcal{N} \gg \sqrt{\frac{\hbar ^2}{Gm^3 l_s}},
\end{equation}
adopt cylindrically symmetric distributions. The corresponding bound on the
string tension is
\begin{equation}  \label{***}
G\mu_{int} \gg \sqrt{\frac{G \hbar ^2}{m l_s \Delta^2}}.
\end{equation}
Thus, even using the Newtonian approximation, we can obtain rough estimates
of the total energy and length scales associated with the BEC string, in both the
interaction and quantum pressure dominated regimes. %

\section{Discussions and final remarks}

\label{Sect.V} We have considered the possible existence, in a cosmological
or astrophysical setting, of static, cylindrically symmetric, general
relativistic structures consisting of matter in a Bose-Einstein condensed
phase or, in other words, of Bose-Einstein condensate strings. By adopting a
semiclassical approach, in which we assume that the quantum dynamics of the
condensate remain unaffected by the gravitational field, and that the
gravitational field remains classical and independent of quantum effects, we
identified two limiting regimes corresponding to thermodynamic (i.e.
particle interaction) pressure and quantum pressure dominated strings. For
both these limiting cases, we solved the gravitational field equations for
the string interior numerically and determined the corresponding variation,
with respect to the radial coordinate, of the components of the metric, the
three-dimensional energy density and pressure, the Tolman mass per unit
length, $M$, and the $W$ parameter which, in conjunction with the initial
conditions for the metric components, determines the angular deficit of the
cylindrical space-time. \newline
\indent
By defining the boundary of the string as the radius at which the energy
density and pressure fall to zero, we also obtained estimates for the order
of magnitude values of the string width, which depend sensitively on the
mass and scattering length of the BEC particles. However, in general, we are
able to conclude that interaction pressure dominates for thick strings,
while quantum pressure dominates for thin strings in which the ratio of
volume to surface area is small. The precise length scale determining the
division between the two is determined by the BEC model parameters. In
principle, both interaction and quantum pressure dominated strings may vary
from tens of kilometers in diameter to widths comparable to more canonical
string species.

Though the space-time of the BEC string exhibits many differences from the
flat conical space-time surrounding a vacuum string (or even the
spherical(-ish) ``cap" that regularizes the vacuum string interior), perhaps
its most interesting feature is that it allows for the existence of an
angle excess: that is, for an angular deficit larger than $2\pi$. Though
this may be considered unphysical, a more intriguing possibility is that
such a string could behave exotically, in the sense of being able to
support a traversable wormhole, as suggested in \cite{Visser1989} - \cite%
{WormholeStrings2}.

The solutions we have obtained describe the interior of the Bose-Einstein
condensate string. The density and pressure both vanish at the string boundary
and therefore, for $r\geq R_s$, an exterior vacuum solution of the
gravitational field equations describes the physical and geometrical
properties of the space-time. Hence, in order to determine the asymptotic form
of the metric, the solutions presented in this paper must be matched onto the
exterior cylindrically symmetric vacuum metric. In the case of
cylindrical symmetry, and by assuming that the exterior gravitational field is described by standard general relativity, this metric is the Kasner metric \cite{Verbin} - \cite{Matt},
\be\label{Kasner}
ds^{2} = (kr)^{2\mathcal{A}}dt^{2} - dr^{2} - \beta^{2} (kr)^{2(\mathcal{B}-1)}r^{2}d{\phi}^2-
(kr)^{2\mathcal{C}}dz^{2} ,
\ee
where $k$ determines the length-scale and $\beta$ is a constant, related to the
deficit angle of the conical spacetime. The Kasner metric is characterized
by two free parameters which, for the unique vacuum solution, satisfy the two Kasner conditions \cite%
{Kasner,exact-sol},
\begin{equation}\label{120}
\mathcal{A} +\mathcal{ B} + \mathcal{C}=\mathcal{A}^2 + \mathcal{B}^2 + \mathcal{C}^2=1.  \label{Kasner2}
\end{equation}

The continuity of the gravitational potentials across the vacuum boundary of the string also imposes the conditions
\bea\label{121}
&&N\left(R_s\right)=\left(kR_s\right)^{\mathcal{A}},L\left(R_s\right)=\beta \left(kR_s\right)^{\mathcal{B}-1}R_s,\nonumber\\
&&K\left(R_s\right)=\left(kR_s\right)^{\mathcal{C}}.
\eea

Equations.~(\ref{120})-(\ref{121}) represent a system of five algebraic equations for the five unknowns ${\mathcal{A}, \mathcal{B}, \mathcal{C},k,\beta}$. Thus, a matching of the interior and exterior solutions would uniquely determine the parameters of the Kasner metric as a function of the physical parameters of the Bose-Einstein condensate forming the string. The standard conic cosmic string solution \cite{clv, Verbin, Verbin1} is characterized by an asymptotic behavior given by a particular form of the Kasner metric (\ref{Kasner}),  with $\mathcal{A} = \mathcal{C} = 0$ and $ \mathcal{B} = 1$. 
In this case, the metric is evidently locally flat, with the parameter $\beta $ representing a conic angular deficit. The direct matching of the Bose-Einstein condensate interior string solution to the locally flat metric requires $N\left(R_s\right)=1$, $K\left(R_s\right)=1$ and $L\left(R_s\right)=\beta R_s$. Due to the initial conditions considered in our numerical analysis of the BEC string model, such a direct matching between the space-time of the massive string interior and a flat exterior geometry is not possible, though this result is consistent with standard results in general relativity \cite{exact-sol}. On the other hand, as pointed out in \cite{Verbin1},  there exists a second interesting Kasner-type solution with $\mathcal{A} = \mathcal{C} = 2/3$ , $\mathcal{B} = -1/3$, which represents the so-called Melvin branch. The Melvin magnetic exterior solution can be matched with the interior BEC string solutions considered in this paper. However, such a matching would require the embedding of the BEC string into an external magnetic field.

Finally, we note that one of the major particle candidates for the formation of BEC strings is dark
matter particles. When the critical temperature of a cosmological boson gas,
which may have existed in the early Universe, became less than the critical
temperature, a Bose-Einstein condensation process may have taken place
during the early cosmic history. Hence, most of the present
day DM may be in the form of a Bose-Einstein condensate.
Thus, during the phase transition from normal to condensate dark matter, BEC-type
topological defects may have been generated, and condensed dark matter
filaments could have been formed. These dark matter filaments may have some
properties in common with the BEC string solutions considered in the present
paper. 

One observational possibility, which may allow us to detect these
structures, and to distinguish them from other string-type species, would be through the detection of gravitational lensing events since, due to their exotic nature,
the lensing properties of BEC strings may differ substantially from those of standard defect, or even superstring candidates (c.f. \cite{topological_defects,Shlaer:2005ry} and references therein). Intriguingly, the possibility of unique, nongravitational lensing phenomena also arises, since some species of non-BEC DM defects are known to lens electromagnetic radiation in a frequency-dependent manner through an alteration of the photon dispersion relation inside the string core, and this lensing is distinct from frequency-independent gravitational lensing \cite{Stadnik:2014cea}. Further investigation of both the gravitational and nongravitational lensing properties of BEC strings may therefore be extremely fruitful.

On the other hand, strings (i.e. topological defects) were previously considered as possible
seeds for structure formation, but this picture has since been overturned in
favor of dark matter seeds. As such, BEC DM strings which represent, in some sense, a
combination of strings and dark matter, may turn out to be a better solution
for large scale structure formation than either nonstringy DM or non-DM strings
individually. In particular, galaxy formation from strong primordial inhomogeneities, such as archioles or clouds of primordial black holes concentrated around intermediate mass, or supermassive black holes could provide a specific implementation of Bose-Einstein condensate strings \cite{Sakharov:1994id,Sakharov:1996xg,Khlopov:1999tm,Khlopov:1998uj,Rubin:2001yw,Khlopov:2002yi,Khlopov:2004sc,Khlopov:2004rw,Khlopov_meeting,Khlopov:2008qy}. 
In the present paper we have provided some basic theoretical
tools that would enable the in depth investigation of the properties of the
BEC strings, and of their cosmological implications.


\section*{Acknowledgments}

We thank Maxim Khlopov and Yevgeny Stadnik for helpful comments during our revision of the manuscript. T. H. thanks the Department of Physics of the Sun-Yat Sen
University in Guangzhou, People's Republic of China, for the kind hospitality offered during the
preparation of this work. %

%


\begin{thebibliography}{99}
\bibitem{exp} M. H. Anderson, J. R. Ensher, M. R. Matthews, C. E. Wieman, and
E. A. Cornell, Science \textbf{269}, 198 (1995); C. C. Bradley, C. A.
Sackett, J. J. Tollett, and R. G. Hulet, Phys. Rev. Lett. {\bf 75}, 1687 (1995); K. B.
Davis, M. O. Mewes, M. R. Andrews, N. J. van Druten, D. S. Durfee, D. M.
Kurn, and W. Ketterle, Phys. Rev. Lett. {\bf 75}, 3969 (1995).

\bibitem{Da99} F. Dalfovo, S. Giorgini, L. P. Pitaevskii, and S. Stringari,
Rev. Mod. Phys. \textbf{71}, 463 (1999).

\bibitem{rev} E. A. Cornell and C. E. Wieman, Rev. Mod. Phys. \textbf{74},
875 (2002); W. Ketterle, Rev. Mod. Phys. \textbf{74}, 1131 (2002); R. A.
Duine and H. T. C. Stoof, Phys. Rep. \textbf{396}, 115 (2004).

\bibitem{Ch05} Q. Chen, J. Stajic, S. Tan and K. Levin, Phys. Rep. \textbf{%
412}, 1 (2005).

\bibitem{Pit} L. Pitaevskii and S. Stringari, \textit{Bose-Einstein
Condensation} (Clarendon Press, Oxford, 2003).

\bibitem{Pet} C. J. Pethick and H. Smith, \textit{Bose-Einstein
Condensation in Dilute Gases} (Cambridge University Press, 2008).

\bibitem{Zar} A. Griffin, T. Nikuni, and E. Zaremba, \textit{Bose-condensed
Gases at Finite Temperatures} (Cambridge University Press, 2009).

\bibitem{Sin} S. J. Sin, Phys. Rev. D\textbf{50}, 3650 (1994); S. U. Ji and
S. J. Sin, Phys. Rev. D\textbf{50}, 3655 (1994); W. Hu, R. Barkana, and A.
Gruzinov, Phys. Rev. Lett. \textbf{85}, 1158, (2000); J. Goodman, New
Astron. \textbf{5}, 103 (2000); P. J. E. Peebles, Astrophys. J. \textbf{534%
}, L127 (2000); A. Arbey, J. Lesgourgues and P. Salati, Phys. Rev. D\textbf{
68}, 023511 (2003).

\bibitem{SF-BEC} E. Castellanos, C. Escamilla-Rivera, A. Mac'as and D. N{\' u%
}{\~ n}ez, J. Cosmol. Astropart. Phys. 11 (2014) 034; A. Su{\' a}rez, V. Robles and T. Matos,
Astrophys. Space Sci. Proc. {\bf 38}, 9, p. 107 (2013); K. Kirsten and D.J. Toms,
Phys. Rev. D \textbf{51}, 6886 (1995); B.-H.
Li, 
Ph.D. Thesis, University of Texas, 2013.

\bibitem{BoHa07} C. G. Boehmer and T. Harko, J. Cosmol. Astropart. Phys. 06 (2007) 025.

\bibitem{inv} J.-W.Lee, Phys. Lett. B \textbf{681}, 118 (2009); J.-W. Lee
and S. Lim,  J. Cosmol. Astropart. Phys. \textbf{01} (2010) 007; T. Harko,  J. Cosmol. Astropart. Phys. 05,
(2011) 022; V. H. Robles and T. Matos, Mon. Not. R. Astron. Soc. \textbf{%
422}, 282 (2012); M. Dwornik, Z. Keresztes, and L. A. Gergely, in
{\it Recent Development in Dark Matter Research}, edited by N. Kinjo, A. Nakajima,
(Nova Science Publishers, New York, 2014); F. S. Guzman, F. D.
Lora-Clavijo, J. J. Gonzalez-Aviles, and F. J. Rivera-Paleo, Phys. Rev. D {\bf 89}, 063507 (2014);
(2013); T. Harko and E. J. M. Madarassy, J. Cosmol. Astropart. Phys. \textbf{01} (2012) 020; B.
Kain and H. Y. Ling, Phys. Rev. D \textbf{82}, 064042 (2010); N. T. Zinner,
Physics Research International, 734543 (2011); P.-H. Chavanis,
Phys. Rev. D \textbf{84}, 043531 (2011); P.-H. Chavanis and L. Delfini,
Phys. Rev. D \textbf{84}, 043532 (2011); P.-H. Chavanis, Phys. Rev. D \textbf{%
84}, 063518 (2011); P.-H. Chavanis, Phys. Rev. E \textbf{84}, 031101
(2011); T. Rindler-Daller and P. R. Shapiro, Mon. Not. R. Astron. Soc.
\textbf{422}, 135 (2012); T. Rindler-Daller and P. R. Shapiro, Mod. Phys. Lett. A {\bf 29}, 1430002 (2014); 
M. O. C. Pires and J. C. C. de Souza, J. Cosmol. Astropart.  Phys. \textbf{11%
}, 024 (2012); T. Harko, Phys. Rev. D \textbf{83}, 123515 (2011); T. Harko
and G. Mocanu, Phys. Rev. D \textbf{85}, 084012 (2012); T. Harko, Mon. Not.
R. Astron. Soc. \textbf{413}, 3095 (2011); P.-H. Chavanis, Astron.
Astrophys. \textbf{537}, A 127 (2012); R. C. Freitas and S. V. B. Goncalves,
J. Cosmol. Astropart. Phys.  \textbf{04} (2013) 049; H. Velten and E. Wamba, Phys. Lett. B \textbf{%
709}, 1 (2012); E. J. M. Madarassy and V. T. Toth, Comput. Phys.
Commun. \textbf{184}, 1339 (2013); F. S. Guzman, F. D. Lora-Clavijo,
J. J. Gonzalez-Aviles, and F. J. Rivera-Paleo, J. Cosmol. Astropart. Phys.  \textbf{09} (2013) 034;
J. C. C. de Souza and M. O. C. Pires, J. Cosmol. Astropart. Phys. {\bf 03} (2014) 010; V. T. Toth,
arXiv:1402.0600; T. Harko, Phys. Rev. D \textbf{89}, 084040 (2014);
M.-H. Li and Z.-B. Li, Phys. Rev. D \textbf{89}, 103512 (2014).

\bibitem{BCS-BEC} A. Morris, http://www.tcm.phy.cam.ac.uk/$\sim$ndd21/%
%

\bibitem{csc} M. G. Alford, K. Rajagopal, T. Schaefer and A. Schmitt,
Rev. Mod. Phys. {\bf 80}, 1455 (2008); M.
Randeria and E. Taylor, 
Ann. Rev. Condensed Matter Phys. {\bf 5}, 209 (2014).

\bibitem{Bal95} M. Baldo, U. Lombardo, and P. Schuck, \prc {\bf 52}, 975
(1995); H. Stein, A. Schnell, T. Alm, and G. Ropke, Z. Phys. A \textbf{351},
295 (1995); U. Lombardo, P. Nozieres, P. Schuck, H.-J. Schulze, and A.
Sedrakian, \prc {\bf 64}, 064314 (2001); E. Babaev, \prb {\bf 63}, 184514
(2001); B. Kerbikov, Phys. At. Nucl. \textbf{65}, 1918 (2002); P.
Castorina, G. Nardulli, and D. Zappala, \prd {\bf 72}, 076006 (2005); A. H.
Rezaeian and H. J. Pirner, Nucl. Phys. A \textbf{779}, 197 (2006); J. Deng,
A. Schmitt, and Q. Wang, Phys. Rev. D \textbf{76}, 034013 (2007); T.
Brauner, Phys. Rev. D \textbf{77}, 096006 (2008); M. Matsuzaki, Phys. Rev. D
\textbf{82}, 016005 (2010); H. Abuki, G. Baym, T. Hatsuda, and N. Yamamoto, %
\prd {\bf 81} 125010 (2010).

\bibitem{NiAb05} Y. Nishida and H. Abuki, \prd {\bf 72}, 096004 (2005).

\bibitem{Sun:2007fc} 
  G.~f.~Sun, L.~He and P.~Zhuang,
  Phys.\ Rev.\ D {\bf 75}, 096004 (2007)
  [hep-ph/0703159].

\bibitem{He:2007kd} 
  L.~He and P.~Zhuang,
  Phys.\ Rev.\ D {\bf 75}, 096003 (2007)
  [hep-ph/0703042].

\bibitem{He:2007yj} 
  L.~He and P.~Zhuang,
  Phys.\ Rev.\ D {\bf 76}, 056003 (2007)
  [arXiv:0705.1634 [hep-ph]].

\bibitem{He:2010nb} 
  L.~He,
  Phys.\ Rev.\ D {\bf 82}, 096003 (2010)
  [arXiv:1007.1920 [hep-ph]].
  
\bibitem{He:2013gga} 
  L.~He, S.~Mao and P.~Zhuang,
  Int.\ J.\ Mod.\ Phys.\ A {\bf 28}, 1330054 (2013)
  [arXiv:1311.6704 [cond-mat.quant-gas]].
  %

\bibitem{Gl00} N. K. Glendenning, {\it Compact Stars, Nuclear Physics, Particle
Physics and General Relativity} (Springer, New York, 2000).

\bibitem{BaBa03} S. Banik and D. Bandyopadhyay, \prd {\bf 67}, 123003 (2003).

\bibitem{Ban04} S. Banik, M. Hanauske, D. Bandyopadhyay, and W. Greiner, %
\prd {\bf 70}, 123004 (2004).

\bibitem{LambdaHyperon} S. Weissenborn, D. Chatterjee and J.
Schaffner-Bielich,
Nucl. Phys. A \textbf{881}, 62-77 (2012);
S. Weissenborn, D. Chatterjee and J. Schaffner-Bielich,
Phys. Rev. C \textbf{85}, 065802 (2012).

\bibitem{Hdibar} T. Sakaia, K. Yazaki and K. Shimizu,
Nucl. Phys. A \textbf{594}, 247 (1995); T. Sakai, K. Shimizu and K. Yazaki,
Prog. Theor. Phys. Suppl. \textbf{137}, 121 (2000); N. K. Glendenning and J. Schaffner-Bielich,
http://ie.lbl.gov/nsd1998/theory/.
%

\bibitem{Ka04} J. I. Kapusta, \prl {\bf 93}, 251801 (2004).

\bibitem{Ab06} H. Abuki, Nucl. Phys. A \textbf{791}, 117 (2007).

\bibitem{ChHa} P. H. Chavanis and T. Harko, Phys. Rev. D \textbf{86}, 064011
(2012); P. H. Chavanis, arXiv:1412.0005 (2014); S. Latifah, A. Sulaksono,
and T. Mart, Phys. Rev. D \textbf{90}, 127501 (2014).

\bibitem{Ba01} C. Barcelo, L. Liberati, and  M. Visser, Classical Quantum
Gravity \textbf{18}, 1137 (2001).

\bibitem{Kolo} E. B. Kolomeisky, T. J. Newman, J. P. Straley, and X. Qi,
Phys. Rev. Lett. \textbf{85}, 1146 (2000).

\bibitem{Ostriker(1964)}  J. Ostriker, Astrophys. J. {\bf 140}, 1056 (1964).

\bibitem{Schneider+Schmitz(1995)}
M. Schneider and F. Schmitz,   Astron. Astrophys. {\bf 301}, 933 (1995).

\bibitem{Christodoulou+Kazanas(2007)} 
D. M. Christodoulou, and D. Kazanas, arXiv:0706.3205 [Astron. Astrophys. (to be published)].
%
%

\bibitem{Carl} S. Carlip, Classical Quantum
Gravity \textbf{25}, 154010 (2008).

\bibitem{clv} M. Christensen, A.L. Larsen and Y. Verbin, Phys. Rev. D
\textbf{60}, 125012 (1999).

\bibitem{Anderson} M. R. Anderson, \textit{The Mathematical Theory of Cosmic
Strings: Cosmic Strings in the Wire Approximation} (Institute of Physics
Publishing, Bristol, 2003).

\bibitem{topological_defects} A. Vilenkin and E.S. Shellard, \textit{Cosmic
Strings and other Topological Defects} (Cambridge University Press, 2000).
%
%

\bibitem{NambuGoto} T. Goto, Prog. Theor. Phys. \textbf{46}, 1560 (1971).
Y. Nambu, Nucl. Phys. B \textbf{130}, 505 (1977).
%

\bibitem{no} H.~B.~Nielsen and P.~Olesen,
Nucl.\ Phys.\ B \textbf{61}, 45 (1973). 
%

\bibitem{NiOl87} N.~K.~Nielsen and P.~Olesen,
Nucl. Phys. B \textbf{291}, 829 (1987).

\bibitem{BlOlVi01} J.~J.~Blanco-Pillado, K.~D.~Olum and A.~Vilenkin,
Phys. Rev. D \textbf{63}, 103513 (2001) [astro-ph/0004410] [SPIRES].

\bibitem{CoHiTu87} E.~J.~Copeland, M.~Hindmarsh and N.~Turok,
Phys. Rev. Lett. \textbf{58}, 1910 (1987). %
%

\bibitem{Anderson et al.(1997)} M. Anderson, F. Bonjour, R. Gregory and J.
Stewart, Phys. Rev. D \textbf{56}, 8014 (1997).
%

\bibitem{Gregory(1988)} R. Gregory, \textit{Yale Cosmic String Workshop}, Yale University, New Haven, 1988 (World Scientific, Singapore, 1988). 
%

\bibitem{Gregory(1988)*} R. Gregory, Phys. Lett, B \textbf{206}, 199 (1988).
%
%

\bibitem{HarkoLake2014} T. Harko and M. Lake, 
(to be published).

\bibitem{MaedaTurok(1988)} K.-I. Maeda and N. Turok, Phys. Lett. B \textbf{%
202}, 376 (1988).

\bibitem{LaLi} L. D. Landau and E. M. Lifshitz, {\it The Classical Theory of
Fields} (Butterworth-Heinemann, Oxford, 1998).

\bibitem{Verbin} Y. Verbin,
Phys. Rev. D \textbf{59}, 105015 (1999). 

\bibitem{Kasner} E. Kasner, Am. J. Math. \textbf{43}, 217 (1921);
E. Kasner, Trans. Am. Math. Soc. \textbf{27}, 155 (1925).
%

\bibitem{Harvey} A. Harvey, Gen. Relativ. Gravit. \textbf{22}, 1433 (1990).
%

\bibitem{exact-sol} D. Kramer, H. Stephani, E. Herlt and M. MacCallum,
\textit{Exact Solutions of Einstein's Field Equations} (Cambridge University
Press, 1980).

\bibitem{Visser1989} M. Visser,
Phys. Rev. D \textbf{39}, 3182 (1989).

\bibitem{WormholeStrings1} G. Cl{\'e}ment,
Ann. Phys. (N.Y.) {\bf 201}, 241 (1990); G. Cl{\'e}ment and I. Zouzou, Phys. Rev.
D \textbf{50}, 7271 (1994).

\bibitem{WormholeStrings2} J. G. Cramer \textit{et al},
Phys. Rev. D \textbf{51}, 3117 (1995); 
G. Cl{\'e}ment, 
Phys. Rev. D \textbf{51}, 6803 (1995); 
C. Barcel and M. Visser,
Nucl. Phys. B \textbf{584}, 415 (2000). 

\bibitem{axion} K. Blum, R. Tito D'Agnolo, M. Lisanti, and B. R. Safdi,
Phys. Lett. B \textbf{737} 30 (2014).


\bibitem{Sikivie:2009qn} 
  P.~Sikivie and Q.~Yang,
  Phys.\ Rev.\ Lett.\  {\bf 103}, 111301 (2009).
  
\bibitem{Budker:2013hfa} 
  D.~Budker, P.~W.~Graham, M.~Ledbetter, S.~Rajendran and A.~O.~Sushkov,
  Phys.\ Rev.\ X {\bf 4}, 021030 (2014).

\bibitem{Roberts:2014dda} 
  B.~M.~Roberts, Y.~V.~Stadnik, V.~A.~Dzuba, V.~V.~Flambaum, N.~Leefer and D.~Budker,
  Phys.\ Rev.\ Lett.\  {\bf 113}, 081601 (2014).

\bibitem{Valent} E. Di Valentino, E. Giusarma, M. Lattanzi, A. Melchiorri,
and O. Mena, Phys. Rev. D \textbf{90}, 043534 (2014).

\bibitem{Verbin1} Y. Verbin, S. Madsen, A. L. Larsen, and M. Christensen,  Phys. Rev. {\bf D 65},  063503 (2002).

\bibitem{Matt} T. Harko and M. J. Lake, arXiv:1409.8454. 

\bibitem{Shlaer:2005ry} 
  B.~Shlaer and M.~Wyman,
  Phys.\ Rev.\ D {\bf 72}, 123504 (2005).
  
  
\bibitem{Stadnik:2014cea} 
  Y.~V.~Stadnik and V.~V.~Flambaum,
  Phys.\ Rev.\ Lett.\  {\bf 113}, no. 15, 151301 (2014).


\bibitem{Sakharov:1994id} 
  A.~S.~Sakharov and M.~Y.~Khlopov,
  Phys.\ Atom.\ Nucl.\  {\bf 57}, 485 (1994)
  
\bibitem{Sakharov:1996xg} 
  A.~S.~Sakharov, D.~D.~Sokoloff and M.~Y.~Khlopov,
  Phys.\ Atom.\ Nucl.\  {\bf 59}, 1005 (1996)
  
\bibitem{Khlopov:1999tm} 
  M.~Yu.~Khlopov, A.~S.~Sakharov and D.~D.~Sokoloff,
  Nucl.\ Phys.\ Proc.\ Suppl.\  {\bf 72}, 105 (1999).
  
\bibitem{Khlopov:1998uj} 
  M.~Yu.~Khlopov, A.~S.~Sakharov and D.~D.~Sokoloff,
 [hep-ph/9812286].
  
\bibitem{Rubin:2001yw} 
  S.~G.~Rubin, A.~S.~Sakharov and M.~Yu.~Khlopov,
  J.\ Exp.\ Theor.\ Phys.\  {\bf 92}, 921 (2001)
  
\bibitem{Khlopov:2002yi} 
  M.~Yu.~Khlopov, S.~G.~Rubin and A.~S.~Sakharov,
  [astro-ph/0202505].

\bibitem{Khlopov:2004sc} 
  M.~Yu.~Khlopov, S.~G.~Rubin and A.~S.~Sakharov,
  Astropart.\ Phys.\  {\bf 23}, 265 (2005).
  
\bibitem{Khlopov:2004rw} 
  M.~Yu.~Khlopov and S.~G.~Rubin, {\it Cosmological pattern of microphysics in the inflationary universe, Fundamental Theories of Physics} (Kluwer, Dordrecht, 2004), Vol. 144.
  
\bibitem{Khlopov_meeting} M. Yu. Khlopov
J. of Phys. Conf. Ser. {\bf 66}, 012032 (2007); in  {\it XXIX
Spanish relativity meeting, Palma de Mallorca, 2007} (IOP, Bristol, 2007)

\bibitem{Khlopov:2008qy} 
  M.~Yu.~Khlopov,
  Res.\ Astron.\ Astrophys.\  {\bf 10}, 495 (2010).

  

\end{thebibliography}
\end{document}